\documentclass[10pt]{iopart}

\usepackage{graphicx} 
\usepackage{subfig}
\usepackage{romannum}
\usepackage{verbatim}
\usepackage{iopams}
\expandafter\let\csname equation*\endcsname\relax

\expandafter\let\csname endequation*\endcsname\relax

\usepackage{amsmath}

\usepackage{epstopdf}
\usepackage{amsfonts}
\usepackage{pifont}
\usepackage{pgf}
\usepackage{pgffor}
\usepackage{subfig}
\usepackage{hyperref}
\usepackage{enumerate}
\usepackage[sort&compress, numbers]{natbib}

\newcommand{\msh}[1]{{\textbf{\color{red}{[Revise: #1]}}}}

\newcommand{\twopiL}{\frac{2\pi}{L}}
\newcommand{\RR}{\mathbb{R}}
\newcommand{\CC}{\mathbb{C}}

\newcommand{\uinf}{\vec{U}_\infty}
\newcommand{\uwake}{\vec{U}_\text{wake}}
\newcommand{\upan}{\vec{U}_\text{panel}}
\newcommand{\ucoll}{\vec{U}_\text{coll}}
\newcommand{\pangam}{\Gamma^\text{panel}}
\newcommand{\npan}{P}
\newcommand{\pannorm}{\hat{n}}
\newcommand{\uprime}{U'}

\graphicspath{{./figures/}}

\begin{document}

\title{Detecting exotic wakes with hydrodynamic sensors}

\author{Mengying Wang and Maziar S. Hemati}

\address{Aerospace Engineering \& Mechanics, University of Minnesota, Minneapolis, MN 55455, USA}
\ead{mhemati@umn.edu}
\begin{abstract}

Wake sensing for bioinspired robotic swimmers has been the focus of much investigation
owing to its relevance to locomotion control, especially in the context of schooling and target following.
Many successful wake sensing strategies have been devised based on models of von K\'arm\'an-type wakes;
however, such wake sensing technologies are invalid in the context of exotic 
wake types that commonly arise in swimming locomotion.
Indeed, exotic wakes can exhibit markedly different dynamics, and so must be modeled 
and sensed accordingly.
Here, we propose a general wake detection protocol for distinguishing between wake types 
from measured hydrodynamic signals alone.
An ideal-flow model is formulated and used to demonstrate the general wake detection framework 
in a proof-of-concept study.
We show that wakes with different underlying dynamics impart distinct signatures on a fish-like body,
which can be observed in time-series measurements at a single location on the body surface.
These hydrodynamic wake signatures are used to construct a wake classification library that is then used 
to classify unknown wakes from hydrodynamic signal measurements.
The wake detection protocol is found to have an accuracy rate of over 95\% in the majority of performance studies conducted here.
Thus, exotic wake detection is shown to be viable, which suggests that such technologies 
have the potential to become key enablers of multi-model sensing and 
locomotion control strategies in the future.


\end{abstract}

\maketitle

\section{Introduction}
Marine creatures display a tremendous ability to interact with their hydrodynamic surroundings: 
some are able to orient themselves with respect to incoming currents~\cite{arnoldBR1974}; some 
can skillfully surf wake vortices for locomotive advantage~\cite{liaoSci2003,marrasBES2015}; 
even the blind among them are able to school~\cite{pitcherSci1976} and navigate obstacle-ridden waters~\cite{montgomeryEBF2001,windsorJEB2010_1,windsorJEB2010_2}.
These remarkable feats of situational awareness and hydrodynamic interaction are, in part,
made possible by suitably evolved mechanosensory organs that allow 
these creatures to sense and perceive their hydrodynamic environments.

Hydrodynamic sensing modalities---such as the lateral line systems of fishes, 
the long facial whiskers of harbor seals,
and the arrays of microsetae on the appendages of 
small crustaceans~\cite{bleckmann1994reception,speddingAnnRev2014,triantafyllouAnnRev2016}---enable 
marine creatures to detect spatiotemporal patterns present in the external fluid 
environment. 
%
For example, the lateral line system consists of a collection of superficial 
and canal neuromasts distributed 
across an organism's body, which effectively serve as a set of 
directional-velocity and pressure sensors, respectively~\cite{bleckmann2009lateral}. 
In principle, a creature can infer much information about its surroundings simply by measuring changes in the local flow; 
hydrodynamic signals encode and transmit information about nearby objects and 
flow structures.
%
%
Indeed, hydrodynamic reception allows some marine swimmers to
evade predators or to track prey~\cite{weihsJTB1984,gemmellJRSI2014,pohlmannPNAS2001}.
%
%
%
%

Although it is common to associate hydrodynamic reception as a close-range sensing capability~\cite{windsorJEB2010_2,sichert2009hydrodynamic,bouffanais2010hydrodynamic}, 
the persistence of some hydrodynamic signals in the waters can allow for longer range sensing as well.  
For instance, the vortical wakes of swimming fish persist along the swimming path of a fish,
allowing for wake detection when the relative distance between the signal transmitter and the signal receiver is large (i.e.,~tens of body lengths)~\cite{bleckmann1994reception}. 
In such instances, a passerby would be able to detect the associated hydrodynamic cues and potentially 
discern relevant information about the passerby.
Indeed, the hydrodynamic trails left behind by swimming fish can encode 
a rich amount of information about a particular swimmer and its motion. 
It is commonly suggested that the vortical wakes generated by marine swimmers 
can provide other aquatic animals with useful information about the swimmer's 
size, swimming speed, and even species and sex specific information~\cite{bleckmann1994reception}.
Given the astounding sensory capabilities afforded to 
marine creatures by hydrodynamic sensing and reception,
numerous investigations have explored technological concepts for 
bringing hydrodynamic sensory capabilities to bear in human-engineered systems.
Much effort has focused on developing artificial lateral line systems and 
associated strategies for various sensing and locomotion control tasks 
(see~\cite{triantafyllouAnnRev2016} for an excellent review of the current state of the art).
%
For example, Chambers et al. show that vortex shedding frequency and magnitude 
can be used by underwater vehicles to 
detect a von K\'arm\'an vortex street and navigate 
flows with varying levels of turbulence and unsteadiness~\cite{chambers2014fish}.
In another study, Klein and Bleckman show that having a total of two artificial 
lateral line canals---one on 
either lateral side of the body---is sufficient to discern 
relevant environmental information, such as object position, 
flow speed, and wake shedding frequency~\cite{kleinBJN2011}.

Many investigations have leveraged model-based strategies for 
studying higher-level hydrodynamic perception and locomotion control 
capabilities~\cite{franoschPRL2009,renBB2012,devriesBB2015,colvertJFM2016}.
%
%
%
%
For example, DeVries et al. used an ideal-flow model 
to devise a lateral-line-based flow 
sensing strategy that fuses measurements from 
both velocity and pressure-based sensing 
modalities~\cite{devriesBB2015}. 
Experimental demonstrations of the multi-modal sensing strategy
show that a robot can utilize the fused hydrodynamic 
information for feedback control and successfully 
achieve rheotaxis and station-keeping~\cite{devriesBB2015}.  
In a separate study, Ren and Mohseni used an ideal-flow model 
to demonstrate a von K\'ar\'man wake sensing algorithm~\cite{renBB2012}.  
%
%
In their study, a mathematical model was formulated for the 
flow inside a canal neuromast, which was then 
used to demonstrate the utility of hydrodynamic reception 
for reconstructing the various parameters associated 
with a von K\'arm\'an-type wake.

An ability to sense von K\'arm\'an-type wakes is of practical interest because 
von K\'arm\'an-type wakes arise behind bluff bodies and many biological and 
bioinspired swimmers.
For example, as noted in~\cite{borazjaniJEB2008}, carangiform swimming propulsion leads 
to the shedding of two single
vortices per tail beat, leading to the well-known reverse von K\'arm\'an vortex street.
We note here that von K\'arm\'an-type wakes are a sub-class of ``2S'' wakes---so-called owing to the fact 
that \emph{two single} vortices are shed per cycle~\cite{williamsonJFS1988}.
%
%
%
%
%
%

Despite demonstrated successes of model-based sensing and control strategies, 
it is important to note that the efficacy of 
a model-based strategy is often predicated 
on the reliability of the model at faithfully describing
the relevant hydrodynamic interactions at play.
It stands that  wake sensing strategies developed in the context of von K\'arm\'an-type 2S wakes
are not necessarily justified nor valid in the context of higher-order \emph{exotic wakes}, where 
more than two single vortices are shed per cycle~\cite{arefJFS2006}.
Indeed, exotic wakes are just as relevant as 2S wakes in biological and bioinspired 
swimming~\cite{borazjaniJEB2008,kern2006simulations,lentinkJEB2008,schnipperJFM2009,mooredJFM2012,smits2014}.
For instance, ``eel-like'' anguilliform swimming commonly leads to the shedding of two pairs of 
vortices per tail beat~\cite{borazjaniJEB2008,tytellICB2010}---often called a ``2P'' wake 
owing to the \emph{two pairs} of vortices that are shed per cycle~\cite{williamsonJFS1988}.
Further, 2P wakes have also been associated with wake resonance modes in 
flapping plate experiments~\cite{mooredJFM2012}, providing a physical 
significance of 2P wakes as a sort of ``optimal'' wake type.

The existence and relevance of exotic wakes present a challenge for wake sensing, 
since different wake types can exhibit markedly different dynamics.
This fact suggests that wake sensing strategies would benefit from 
a ``multiple model'' approach, 
in which a wake is first detected and classified by type, 
\emph{then} sensed and reconstructed with a suitable model corresponding to 
the identified wake type.
Indeed, this is one motivation for the exotic wake detection strategy that we introduce and study here.
%
%
%

Beyond the distinctions between 2S, 2P, and higher-order wakes,
dynamical differences can also arise between various sub-classes of these wake types.
A familiar example is the difference between sub-classes of 2S wakes:~i.e., the difference between drag producing 
von K\'arm\'an~(vK) wakes and thrust generating reverse von K\'arm\'an~(rvK) wakes.
Distinct regimes of motion also arise in exotic wakes, 
as has been well established in the vortex dynamics literature~\cite{arefJFS2006,basuPOF2015,stremlerTCFD2010,stremlerJFS2011}.
For instance, in the recent study by Basu and Stremler, 
an idealized point vortex model of a 2P wake was thoroughly examined and 
found to exhibit twelve dynamically distinct regimes of motion~\cite{basuPOF2015}.
%
%
%

%
A careful examination of the biolocomotion liteature suggests that different 
``wake regimes'' (i.e.,~wake exhibiting different dynamical characteristics) are commonly observed for 
a given ``wake type'' (i.e.,~a given wake pattern, such as 2S or 2P), in numerical simulations and physical experiments alike.
%
For instance, Schnipper et al. observe \emph{drag producing} 2P wakes in experiments of a 
flapping foil~\cite{schnipperJFM2009}; whereas, the 2P wakes they aimed to study were 
motivated by the \emph{thrust generating} 2P wakes observed in undulatory swimming~\cite{mullerJEB2008}.
%
%
In numerical studies, Borazjani and Sotiropoulos show that anguilliform swimming can result in 
both drag- and thrust-producing 2P wakes~\cite{borazjaniJEB2008}.
Still more, the gait optimization study of Kern and Koumoutsakos
suggests that an anguilliform swimming gait can be modified 
to achieve different objectives, but will consistently 
generate a 2P wake pattern: e.g.,~swimming to maximize velocity 
and swimming to maximize efficiency both yield 2P wake patterns~\cite{kern2006simulations}.
The dynamical differences between each of these 2P wakes suggests that 
there may be much more information to glean from a wake signature
simply by considering the wake dynamics;
associating a wake signature with a particular wake regime can allow inferences 
about the wake generating system.
%
%
Although future work is still needed to establish connections 
between specific wake dynamics and various swimming characteristics~\cite{hematiAIAA2016},
the study here will demonstrate that different wake regimes impart distinct 
hydrodynamic signatures that can enable wake detection and classification 
from sensor measurements.
For example, a carangiform swimmer (2S wake generator) can be distinguished from 
an anguilliform swimmer (2P wake generator) from hydrodynamic wake 
signatures alone.
Further, it may be possible to determine more refined information by considering 
the sub-regimes of motion, such as whether a particular anguilliform swimmer was 
swimming for speed or efficiency.
In fact, we will show here that wake types and wake regimes can be classified using 
hydrodynamic signals measured at only a single location on the surface of a fish-like body.
To do so, we begin by constructing a library of wake signatures from known wake types. 
Then, given the measured signature from an unknown nearby wake,
the type and dynamical regime of the wake can be determined by comparing 
against entries in the library.
In this study, we explore the viability of classifying wake regimes from 
hydrodynamic signal measurements.
We begin, in Section~\ref{sec:wakedetection}, by introducing a general framework for wake detection.
Then, the remainder of the manuscript will focus on the specific details of 
the proof-of-concept demonstration.
In Section~\ref{sec:modeling}, we present an ideal-flow model 
to represent the dynamics of various wakes
and to study the hydrodynamic signatures that they impart on a fish-like body.
In Section~\ref{sec:results}, we present a detailed study of the wake detection 
strategy in the context of the ideal-flow model. 
Each step in the construction of the wake detection protocol is presented along with  
a performance study.
Finally, in Section~\ref{sec:conclusions}, we discuss the results of the study, draw conclusions, and offer suggestions for future investigations.
\section{A general framework for wake detection and classification}
\label{sec:wakedetection}


The problem of exotic wake detection is concerned with associating measured hydrodynamic signals
with a particular wake type.
Indeed, the exotic wake detection problem can be framed as a problem in time-series classification, which we will address by means of a general supervised learning strategy.
Ultimately, this strategy will allow an unknown wake 
to be detected and classified solely from measured
hydrodynamic signals.
However, before the actual classification task can be accomplished,
the proposed approach will require a library of wake signatures to be constructed.
In constructing this library, it is assumed that the wake types are known, 
such that every entry in the library of wake signatures has a wake type associated with it.
Once the library is constructed, the classification task can be carried out by comparing 
a measured signature from an unknown wake type with the entries in the library.
In this way, the ``closest match'' in the library with the measured signal can be used to 
classify the unknown wake.
Below we outline more specifically the steps that must be taken to construct a library and 
implement a general wake detection and classification protocol.
%
%
%

\begin{figure} [h!]
\begin{center}
\subfloat[Library construction]{
 \includegraphics[width=0.45\textwidth]{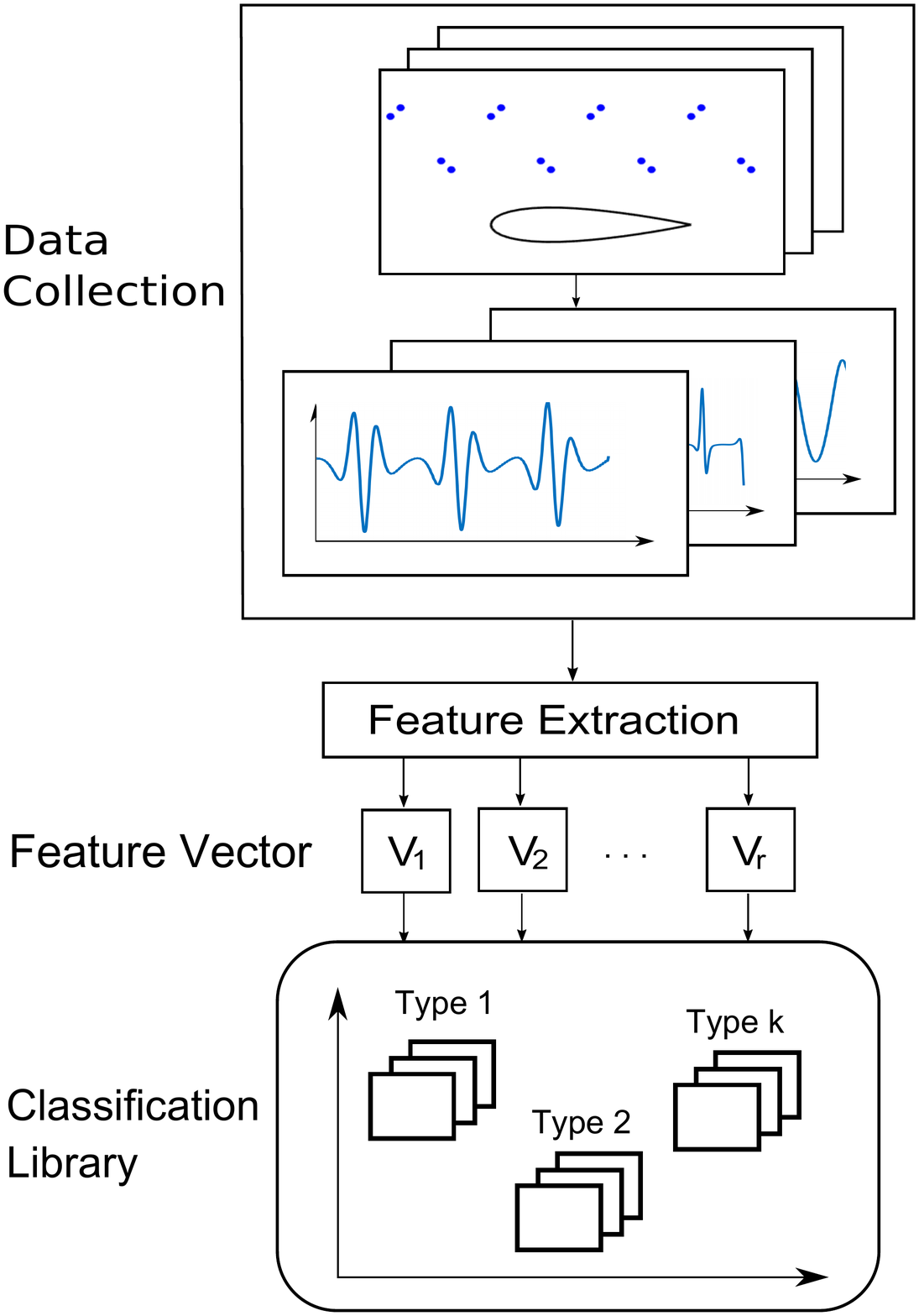}
\label{fig:ConsLib} }
\subfloat[Wake Classification]{
 \includegraphics[width=0.45\textwidth]{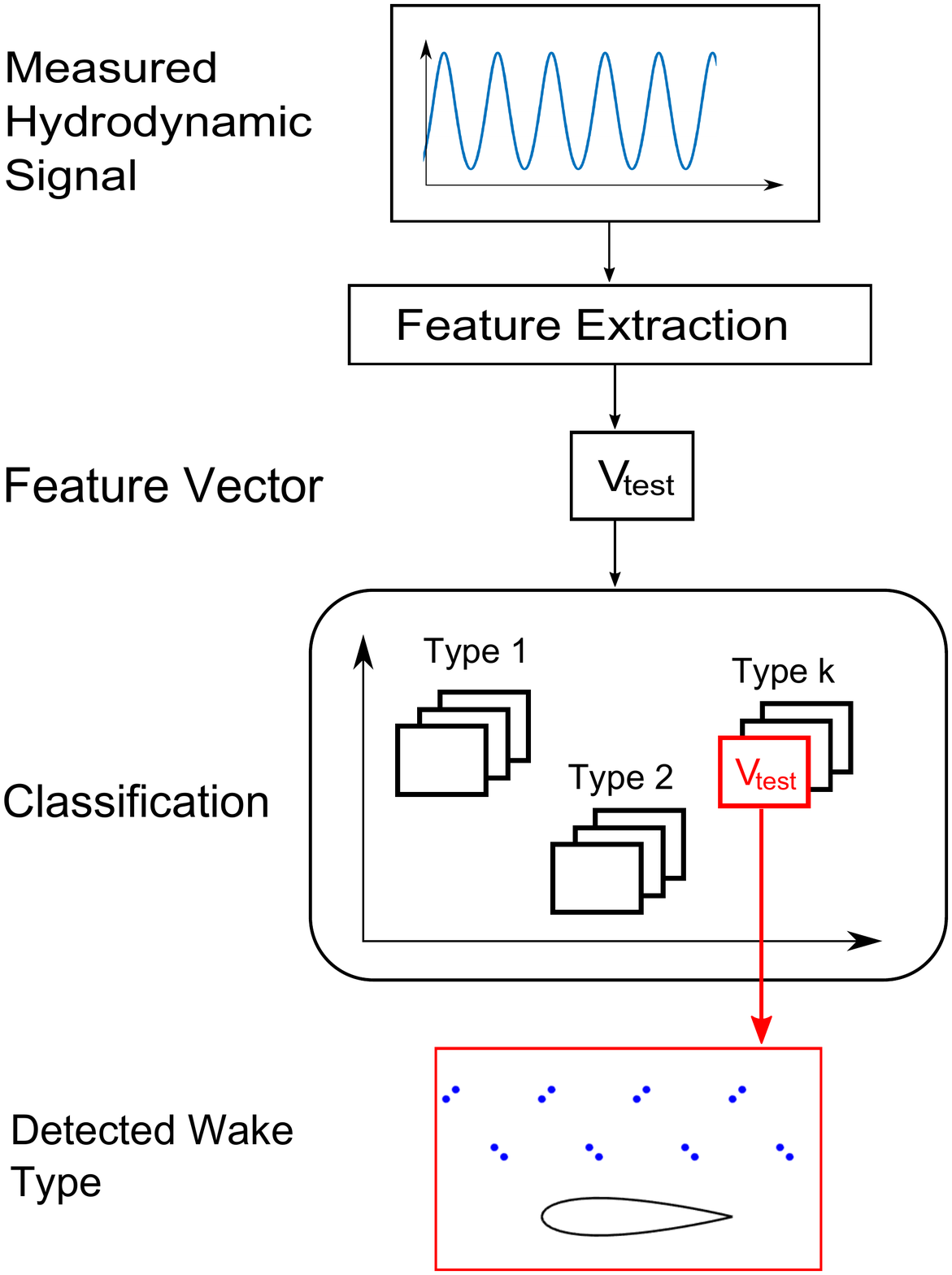}
\label{fig:WakeCla} }
\caption{Synthesis of an exotic wake detection protocol consists of two primary tasks, summarized graphically here. The library construction stage in (a) is performed first, then followed by the wake classification process shown in (b).}
\label{fig:FW1}
\end{center}
\end{figure}

To construct a wake classification library, we begin by collecting sensor measurements over a 
large number of known wake types.
In doing so, it is important to capture multiple realizations from each of the various wake types that are to be included in the library;
this is needed to ensure that the library will be sufficiently rich for the classification task.
Since we are working with time-series data, it will also be useful to perform 
a ``feature extraction'' step that transforms the time-series signal into a ``static'' feature vector, denoted $V_i$, such that
entries in the library will be invariant to
such factors as the start time of data collection.
The primary challenge associated with feature extraction here rests in choosing a feature vector that adequately summarizes
the time-series data and can be used to effectively distinguish between wake types.
With an appropriate feature vector defined,
the feature vectors for all $r$ realizations $\{V_1,V_2,...,V_r\}$ of wake signatures 
can be collected along with associated wake types (i.e.,~``labels''), then stored in a library for 
use during the wake detection and classification task.
The full library construction process is summarized in Figure~\ref{fig:ConsLib}.
%
%

Once the library of wake signatures is constructed, a classification algorithm can be 
applied to classify an unknown wake type from its hydrodynamic signature---called a ``test signal''---by comparing 
with entries in the library. 
%
In order to do so, the time-series data in the test signals must
be converted into a feature vector $V_\text{test}$, as was done in the library construction stage, 
in order to compare with entries in the library.
Classification can then be performed by evaluating which library entries match most closely with $V_\text{test}$.
In the present study, we make use of the $k$-nearest-neighbor (KNN) algorithm to perform this comparison and 
to determine the unknown wake type, primarily owing to its simplicity~\cite{bishop2006,tan2006};
however, this specific algorithm can be replaced by alternative 
classification techniques as well.
%
%
The wake detection and classification procedure is summarized in Figure~\ref{fig:WakeCla}.

In the next section, we present an ideal-flow model that will be used to demonstrate the wake detection and classification approach in Section~\ref{sec:results}.
The ideal-flow model will be used to generate representative wake signature data, which will inform many of the specific implementation 
details that were presented more generally in this section.
Implementation details, such as the choice of feature vectors, will be guided by insights gained from a careful analysis of the specific wake signatures 
that are collected during the library construction processes.

%
%

\section{Theoretical modeling}
\label{sec:modeling}
At this point, we are interested in formulating a modeling framework by which to 
demonstrate the wake detection and classification approach 
described in Section~\ref{sec:wakedetection}.
To this end, we will formulate an ideal-flow model to represent the hydrodynamic 
influence of a dynamic wake on a fish-like body.
We emphasize, however, that the wake detection and classification approach presented in Section~\ref{sec:wakedetection} is 
not model-based; rather, it is entirely data-driven.
The ideal-flow model formulated here is used only to generate 
representative wake signatures for the proof-of-concept studies presented in Section~\ref{sec:results}.
In principle, the wake detection approach can be applied to generate a 
classification library and classify wake signatures obtained in other settings just as well.

Important to the present study is the ability to generate different regimes of 
motion within a given wake type; 
beyond determining whether a wake signature is associated with a 2S or 2P wake,
we will also seek to determine whether a nearby wake is a von K\'arm\'an 2S wake~(vK), 
a reverse von K\'arm\'an 2S wake~(rvK), or any of a number of 2P wake sub-regimes.
%
%
To this end, we represent the ambient wake vorticity as a singly-periodic array of point vortices,
whose dynamics are governed by the associated vortex dynamics equations.
Different wake types can be represented simply by changing the number of point vortices in 
the base-strip, while sub-regimes can be attained by modifying physical 
parameters associated with the vortex system, as will be described in Section~\ref{ssec:wakemodel}.
These singly-periodic wake models are ideally suited 
for the purposes of our study here: 
these models allow various wakes and dynamical sub-regimes to be 
known ahead of time, and thus provide an objective baseline for validating the 
wake detection and classification approach.
%
%
%

Further, we employ a vortex panel method to model the fish-like body,
as will be described in Section~\ref{ssec:vpm}.
%
%
By combining the vortex panel method with the wake dynamics model~(see Figure~\ref{fig:fishwake}),
we are able to model the hydrodynamic signatures of various wake regimes imparted on a fish-like body.
The velocity at a point (or set of points) on the surface of the body 
can then provide time-series measurements that are representative of the 
hydrodynamic signals detected by a superficial neuromast or similar bioinspired velocity sensor.
The modeling approach taken here can be generalized to study multiple sensing modalities.
For example, the canal neuromast model of Ren and Mohseni~\cite{renBB2012}
could be introduced directly---though we do not present 
such results here.
%
%

\subsection{Vortex wake modeling}
\label{ssec:wakemodel}
%
%
Here, we summarize the basic formulation of the wake dynamics
models, then discuss specific cases of 2S and 2P wakes.
%
%
%
The vorticity $\omega(z,t)$, as a function of complex position $z=x+iy$ and time $t$, 
can be represented by a system of $N$ point vortices,
\begin{equation}
  \omega(z,t) = \sum_{\alpha=1}^N\Gamma_\alpha\delta(z-z_\alpha(t)),
\label{eq:vorticity}
\end{equation}
where $\Gamma_\alpha\in\RR$ and $z_\alpha(t)\in\CC$ denote the strength and complex position of vortex $\alpha$, respectively.
Here, we assume that the vortex strengths remain constant in time.
In the unbounded domain, a system of $N$ point vortices evolves according to~\cite{newton2001,cottet2000}
\begin{equation}
\frac{\mathrm{d}{z_\alpha}^* }{\mathrm{d} t} = \frac{1}{2\pi i} \sum_{\beta=1\atop\beta\ne\alpha}^N\frac{\Gamma_\beta}{z_\alpha-z_\beta},
\label{eq:unbounded_eom}
\end{equation}
where ${(\cdot)^*}$ denotes complex-conjugation.

\begin{figure}[h!]
\begin{center}
\subfloat[Idealized 2S wake model $(N = 2)$]{
 \includegraphics[width=0.75\textwidth]{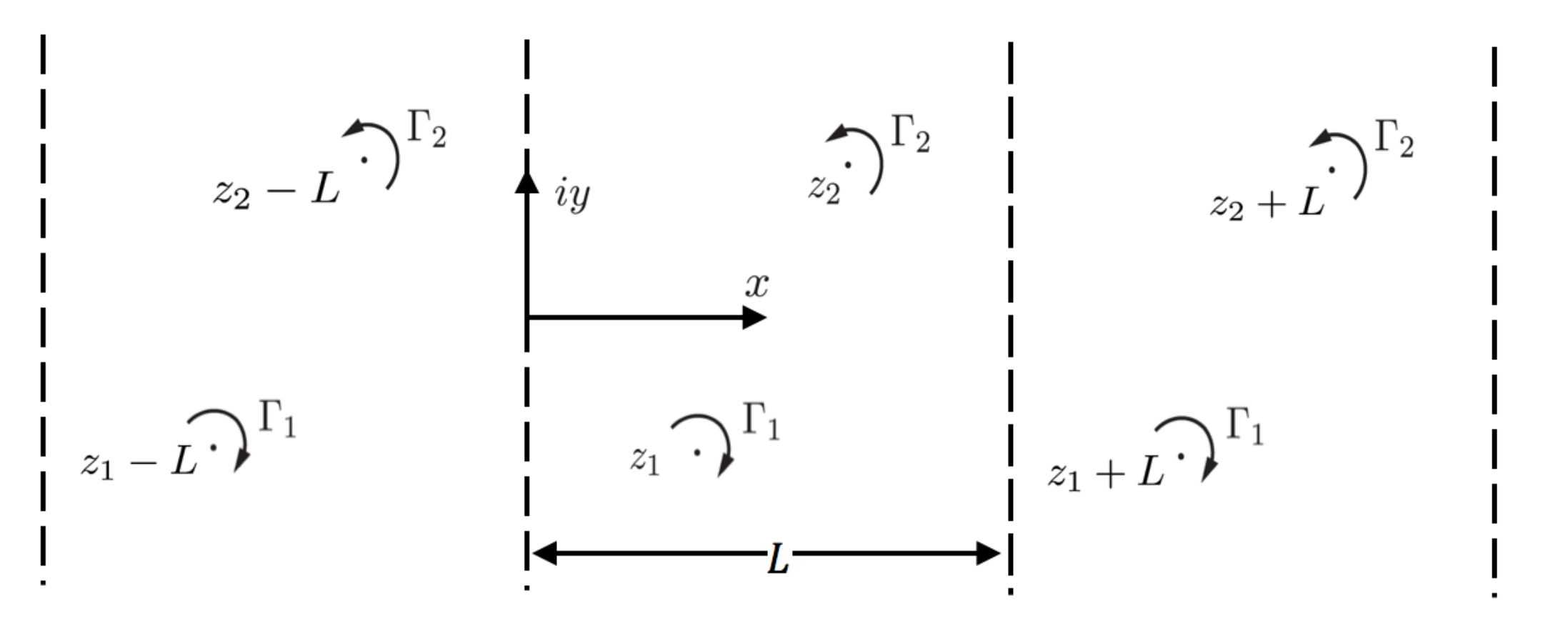}
\label{fig:wake2S} }


\subfloat[Idealized 2P wake model $(N = 4)$]{
  \includegraphics[width=0.75\textwidth]{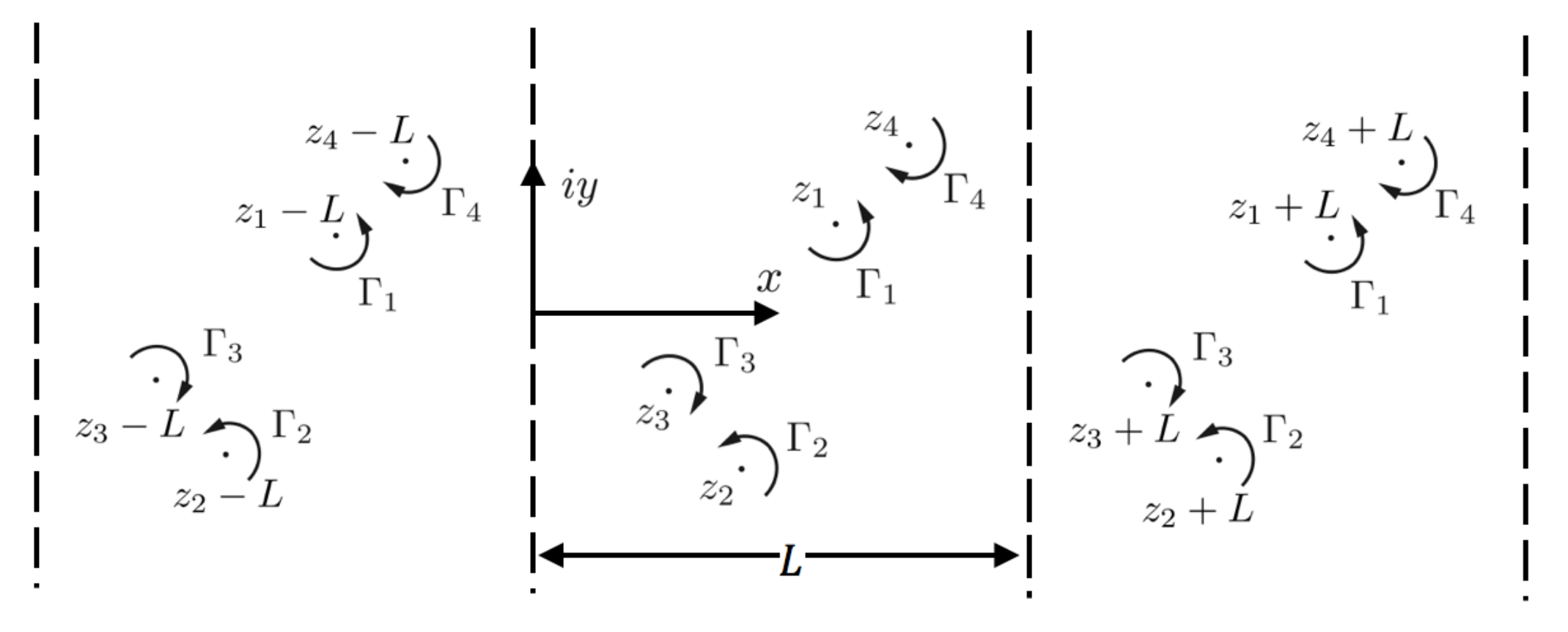}
   \label{fig:wake2P} }
 \caption{The evolution of point vortices on a singly-periodic strip can be used to model wake dynamics.  Here, three spatial periods are plotted for the point vortex models associated with (a)~a 2S wake and (b)~a 2P wake.} \label{fig:wakemodel}
 \end{center}
\end{figure}

In modeling the evolution of periodically shed wake vorticity, we consider the evolution of 
$N$ vortices in a strip of a singly-periodic domain (see~Figure~\ref{fig:wakemodel}).
Thus, we account for the mutual interactions between all the vortices in a given strip (as in Eq.\eqref{eq:unbounded_eom}), as well as the interactions with all of the vortices in the strips along 
the periodic direction (taken as $x$ here).
Taking the length of a single strip to be $L\in\RR$, the equations of motion for the singly-periodic wake reduce to
%
\begin{equation}
  \frac{\mathrm{d} {z_\alpha}^*}{\mathrm{d} t} = \frac{1}{2L i}\sum_{\beta=1\atop \beta\ne\alpha}^N\Gamma_\beta\cot\left[\frac{\pi}{L}(z_\alpha-z_\beta)\right].
\label{eq:vort_eom}
\end{equation}
%
Here, the vortex system is considered from a reference frame that moves with the background flow. 
Since the vortex strengths are assumed to be invariant, the sum of vortex strengths $\Gamma _{\infty }$ is a constant of motion that, in an effort to model the periodic shedding of vorticity into the wake, is taken to be zero:
\begin{equation}
	\Gamma _{\infty } = \sum_{\alpha=1}^{N}\Gamma_\alpha = 0.
\end{equation}

We note that \eqref{eq:vort_eom} can be expressed in Hamiltonian form,
\begin{equation}
  \Gamma_\alpha\frac{\mathrm{d} x_\alpha}{\mathrm{d} t}=\frac{\partial \mathcal{H}}{\partial y_\alpha}, \qquad \Gamma_\alpha\frac{\mathrm{d} y_\alpha}{\mathrm{d} t}=-\frac {\partial \mathcal{H}}{\partial x_\alpha}.
\end{equation}
with the Hamiltonian for the singly-periodic vortex system---a constant of motion---given as,
\begin{equation}
  \mathcal{H}(z_1,\dots,z_N; \Gamma_1,\dots,\Gamma_N) = -\frac{1}{4\pi}\sum_{\alpha=1}^N\sum_{\beta=1\atop \beta\ne\alpha}^N\Gamma_\alpha\Gamma_\beta \ln\left|\sin\left[\frac{\pi}{L}(z_\alpha-z_\beta)\right]\right|. \label{eq:hami}
\end{equation}
Hence, given a set of vortex strengths $\Gamma_\alpha$, the topology of phase-space becomes fixed.
Then, a particular trajectory in phase-space is determined by the initial positions of the base vortices $z_\alpha$.
That is, the vortex strengths fix all possible dynamical regimes, 
while the initial vortex positions determine which among these is actually realized.
%

\subsubsection{2S wake model ($N = 2$)}
We begin by using the point vortex framework described above to model 2S wake dynamics by setting $N=2$ (see Figure~\ref{fig:wake2S}).
Since the equations of motion were formulated assuming zero net circulation in the wake, it follows that the strengths of the two point vortices in each strip are equal and opposite (i.e.,~ ${\Gamma_1 = -\Gamma_2 = \Gamma}$).
The equations of motion for the 2S wake then reduce to
\begin{equation}
  \begin{split}  
    \frac{\mathrm{d} {z_1}^*}{\mathrm{d} t} = -\frac{\Gamma}{2L i}\cot\left[\frac{\pi}{L}(z_1-z_2)\right],\\
    \frac{\mathrm{d} {z_2}^*}{\mathrm{d} t} = \frac{\Gamma}{2L i}\cot\left[\frac{\pi}{L}(z_2-z_1)\right].
  \end{split}
\end{equation}
Here, the complex separation $(z_1-z_2)=(\Delta x + i\Delta y)$ is a constant of motion,
which implies that the motion of wake vortices will be parallel to the wake-axis 
(i.e.,~$-\frac{\Gamma}{2L i}\cot\left[\frac{\pi}{L}(\Delta x+i\Delta y)\right]=0$).
%
%
It follows that the pair of vortices can take on one of two configurations in terms of horizontal spacings: 
(1)~a symmetric wake with $\Delta x=0$, or (2)~a staggered wake with $\Delta x=L/2$.
%
%
For the staggered 2S wake, the vortex configuration is stable when the vertical separation is $\Delta y=\frac{L}{\pi}\sinh^{-1}(1) \approx 0.28L$~\cite{renBB2012}.
Since the staggered configuration corresponds to the von K\'arm\'an-type wakes that are 
commonly observed behind bluff bodies and in swimming locomotion, we will focus on these 
staggered 2S configurations ($\Delta x=L/2,\Delta y=0.28L$) in the remainder of our study here.

\subsubsection{2P wake model ($N = 4$)}
The dynamics of a 2P wake can be modeled by setting $N=4$~(see Figure~\ref{fig:wake2P}).
Here, we only present a summary of these models and their associated dynamics, as needed for the present investigation.
Further details can be found in the recent works by Stremler and colleagues~\cite{basuPOF2015,stremlerTCFD2010,stremlerJFS2011}. 
%
%
%

The 2P vortex dynamics can be reduced to an integrable two-degree-of-freedom 
Hamiltonian system by taking the base vortex strengths and positions to be
\begin{align}
	\Gamma_3=-\Gamma_1,\qquad &  \Gamma_4=-\Gamma_2 \\
    z_3 = z_1^*-\frac{L}{2},\qquad & z_4 = z_2^*+\frac{L}{2}.
\end{align}
Defining $S:=\Gamma_1+\Gamma_2$, it follows that the non-dimensional linear impulse for the base vortices $\mathcal{Q}+i\mathcal{P}=(\pi/LS)\sum_{\alpha=1}^N\Gamma_\alpha z_\alpha$ reduces to
\begin{align}
	\mathcal{Q} &= \frac{\pi}{2}\left(2\gamma-1\right)\\
	\mathcal{P} &= \frac{2\pi}{L}\left[\gamma y_1 + (1-\gamma)y_2\right],
\end{align}
where $\gamma:=\Gamma_1/S$.
For convenience, we take $0\le\Gamma_1\le\Gamma_2$, such that $0\le\gamma\le1/2$.
Then, upon defining $Z:=X+iY=\pi(z_1-z_2)/L$ as the normalized separation between the first and second base-vortices,
the Hamiltonian takes the form~\cite{basuPOF2015}
\begin{equation}
\mathcal{H}=\mathcal{H}(X,Y;\gamma,\mathcal{P}).
\end{equation}
In other words, for the idealized 2P wake model, the evolution of the associated system of point vortices can be
reduced to tracking the separations between the base pairs of vortices.
Further, the motion is parameterized by the linear impulse of the base pairs, though it is more convenient to express this parameterization in terms of the non-dimensional strength $\gamma$ and the vertical component of linear impulse $\mathcal{P}$. 

\begin{figure}[h!]
\begin{minipage}{0.5\textwidth}
	\centering
	\subfloat[]{
    	\includegraphics[width =\textwidth]{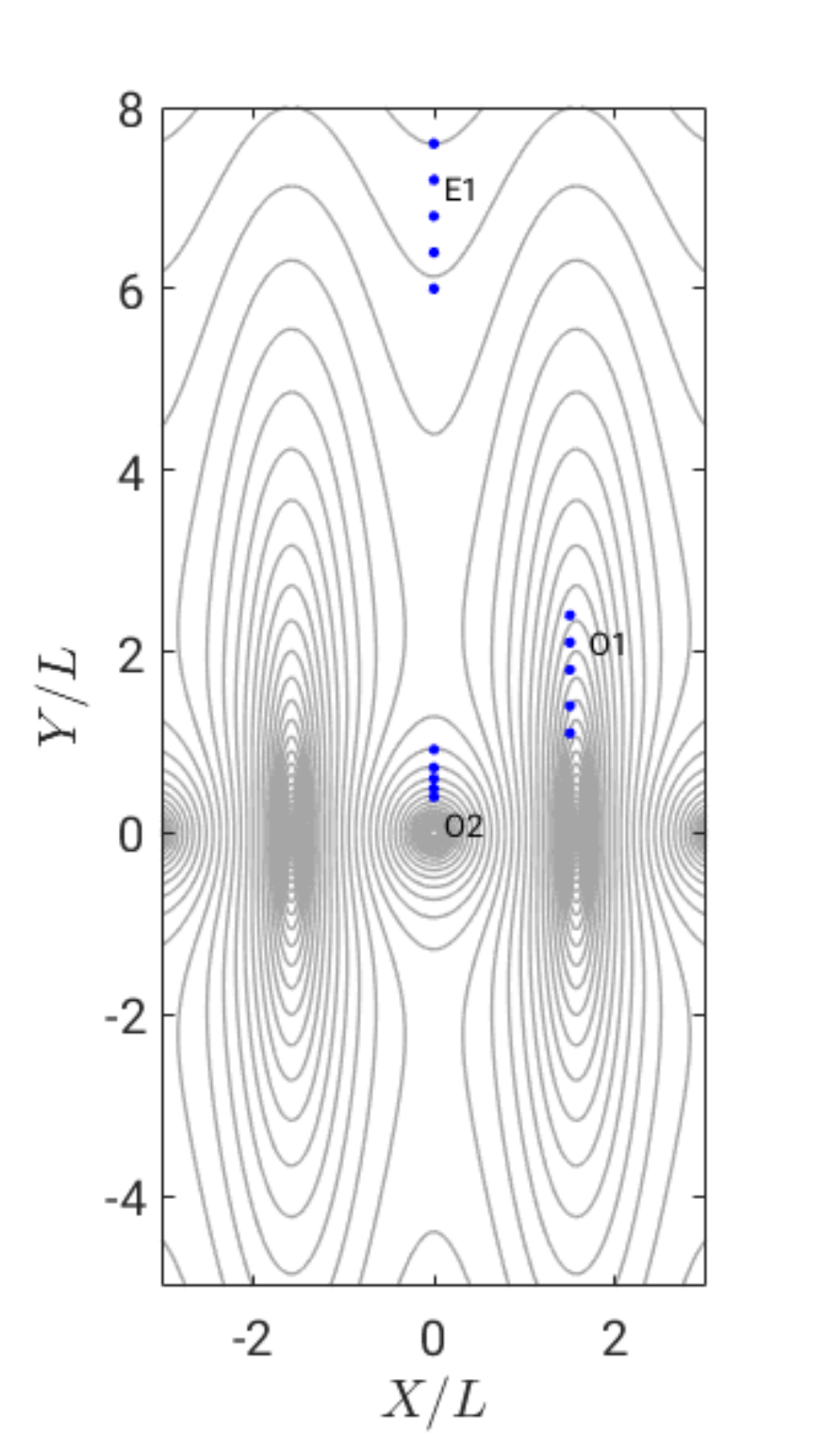}
        \label{fig:hamiX}
    }
   \end{minipage}\hfill
   \begin{minipage}{0.45\textwidth}
   \centering
   \subfloat[E1]{
   		\includegraphics[width=\textwidth]{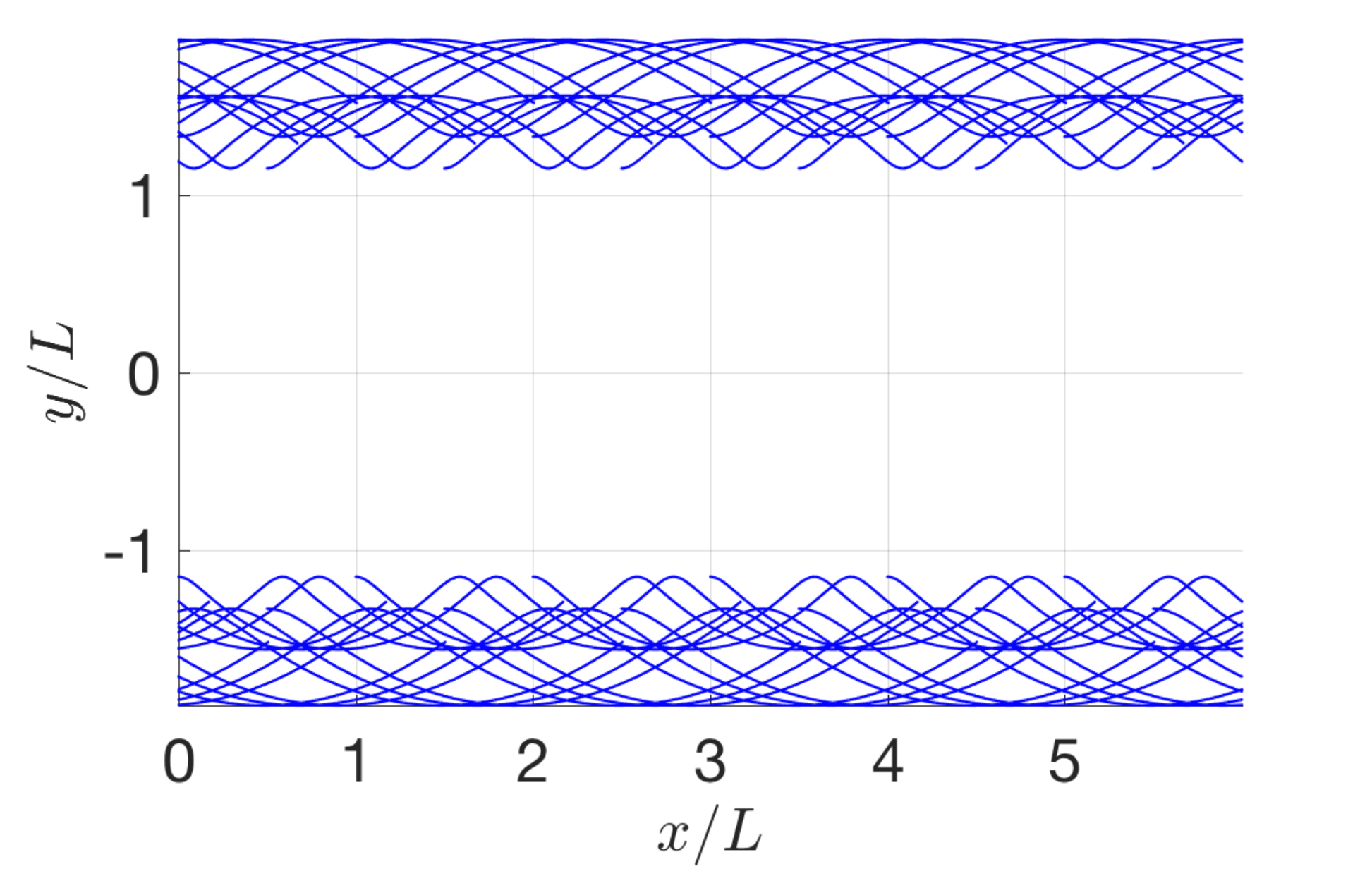}
        \label{figure:traE1}
   }\hfill        
   \subfloat[O1]{
   		\includegraphics[width=\textwidth]{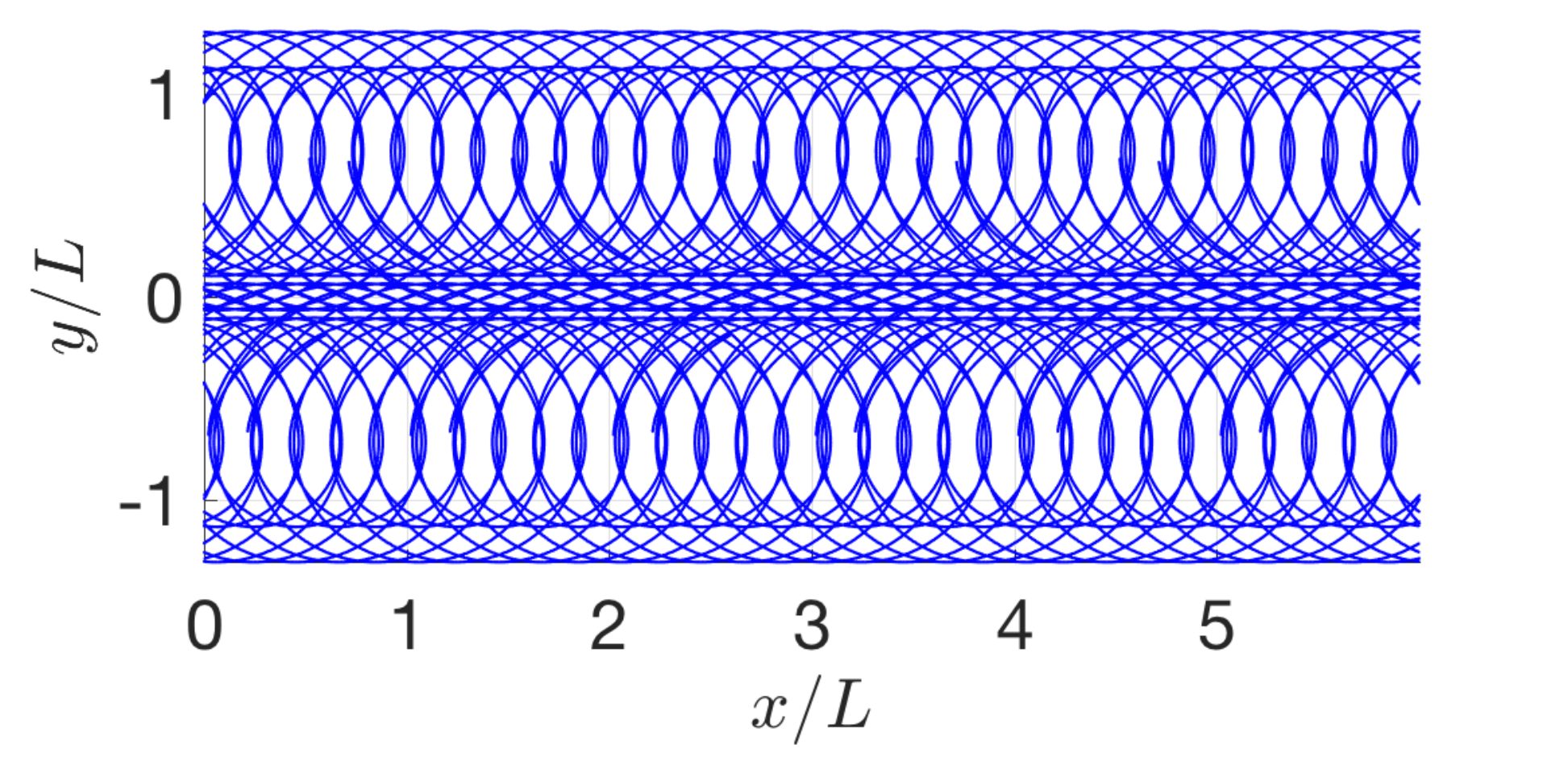}
        \label{figure:traO1}
   }\hfill
   \subfloat[O2]{
   		\includegraphics[width=\textwidth]{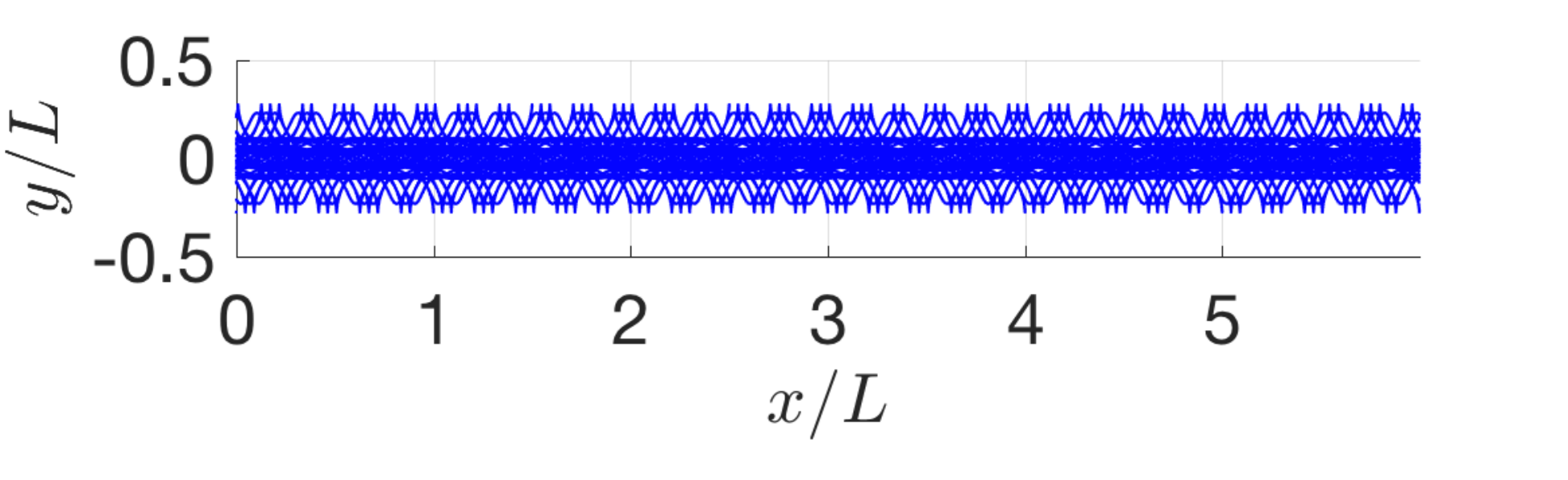}
        \label{figure:traO2}
  }
   \end{minipage}
\caption{Level curves of $\mathcal{H}$ define trajectories with qualitatively different behaviors in phase-space $(X,Y)$ and physical-space $(x,y)$.  (a)~shows level curves of $\mathcal{H}$ corresponding to the 2P wake model with $\gamma = \frac{3}{7}$ and $\mathcal{P} = -0.55$. Three wake regimes are highlighted: $E1$, $O1$, and $O2$.  The blue dots correspond to a sample of initial conditions that result in these classes of motion. (b)~shows a sample trajectory of vortices in physical-space from each of these regimes.}
\label{fig:HamiCurveIX}
\end{figure}        

As shown in~\cite{basuPOF2015}, the idealized 2P wake model can exhibit twelve distinct regimes of motion, 
each characterized by different patterns in the vortex trajectories.
In particular, Basu and Stremler identify numerous sub-classes of orbiting, exchanging, passing, scattering, and mixed patterns in the vortex trajectories.
Figure~\ref{fig:HamiCurveIX} shows level curves of the Hamiltonian for a 2P wake with $\gamma = \frac{3}{7}$ and 
$\mathcal{P} = -0.55$ in tile (a), which corresponds to all possible phase-space trajectories.
We highlight three regimes for this particular wake: an exchanging regime~($E1$) and 
two orbiting regimes~($O1$ and $O2$).
Tiles (b)--(d) in Figure~\ref{fig:HamiCurveIX} show trajectories in physical-space corresponding to each of these regimes.
As is made evident in the figure, $\gamma$ and $\mathcal{P}$ determine all dynamical regimes that are possible, 
while the initial separation $(X,Y)$ (show as blue dots in the figure) determines the value of the 
Hamiltonian $\mathcal{H}$---a constant of motion---and thus dictates which of these regimes is actually realized.
From the figure, it is clear that each wake regime has a distinct pattern that in phase-space and in physical-space.
Our goal in the present investigation will be to distinguish between these patterns from 
corresponding hydrodynamic signal measurements.

As noted in~\cite{basuPOF2015}, the $E1$, $O1$, and $O2$ regimes correspond to ``typical'' motions.
Indeed, unlike other regimes, $E1$, $O1$, and $O2$ arise over a broad range of values for the wake 
parameters $(\gamma,\mathcal{P})$.
Scattering and passing wakes only arise in limiting cases with 
$\gamma = 0$ and $\gamma = 1/2$, respectively.
The mixed wake regime consists of elements of both orbiting and exchanging motions, 
which is expected to be a difficult edge-case and will not be considered here. 
Thus, in the current study, we will only consider the more typical $E1$, $O1$, and $O2$ regimes for 2P wakes,
which will be generated over a range of $(\gamma,\mathcal{P})$ values. 
%
%

\subsection{Fish-like body modeling}
\label{ssec:vpm}
A fish-like body is modeled as a bound vortex sheet by means of 
a vortex panel method~\cite{katz2001}.
Here, the bound vortex sheet is represented as a set of point vortices that are 
appropriately distributed along the body surface.
Further details on vortex panel methods can be found in 
standard texts, such as~\cite{katz2001}.
%
%
%
%

\begin{figure} [h!]
\begin{center}
\includegraphics [width=0.75\textwidth] {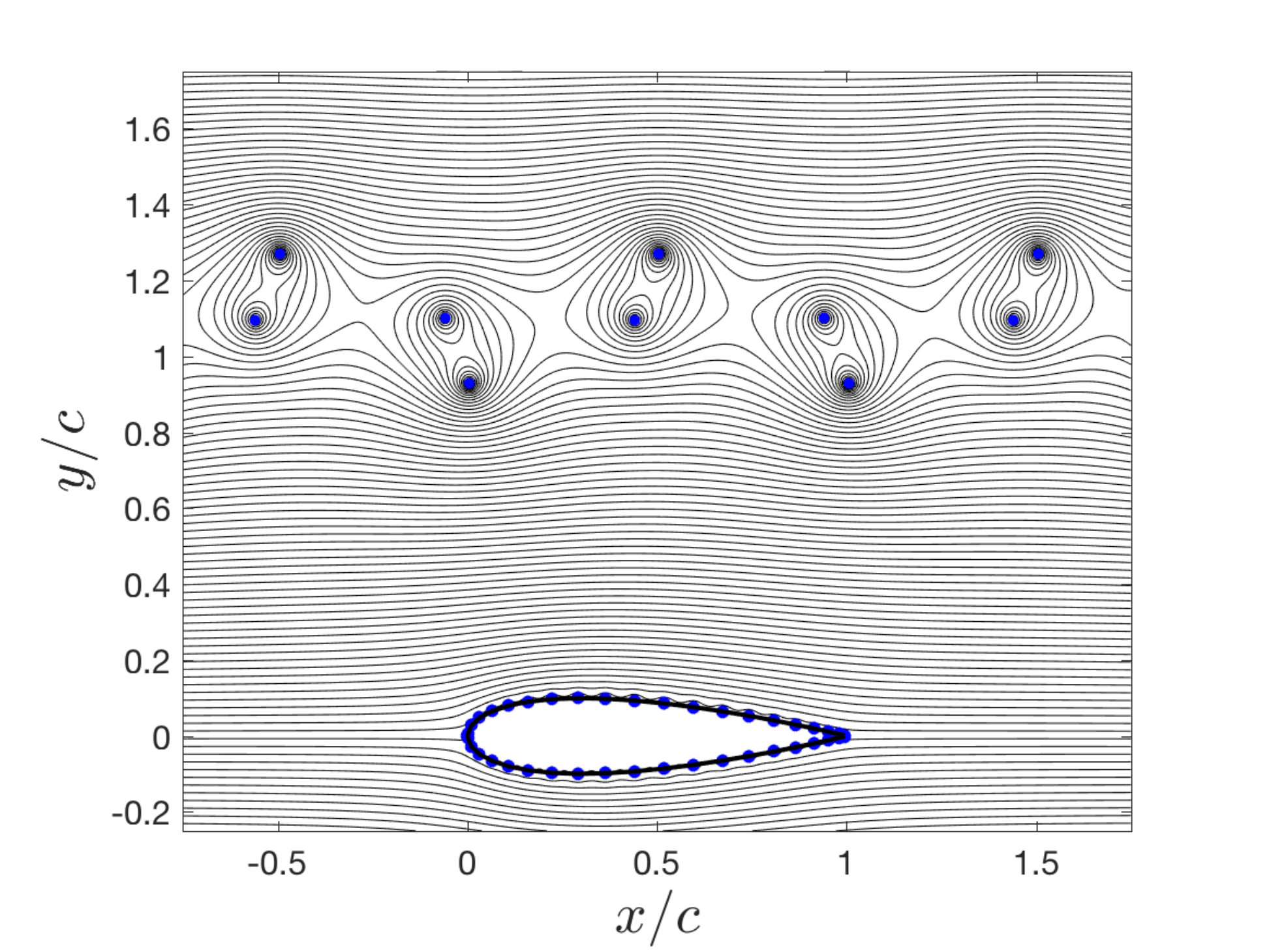}
\caption{A snapshot in the evolution of the 2P wake in the vicinity of a fish-like body.  Free and bound point vortices are depicted as blue dots.  Thick black contour outlines the geometry of the NACA~0020 airfoil.  The no-flow-through boundary condition at the panel collocation points is respected, as seen in the streamlines.}
\label{fig:fishwake}
\end{center}
\end{figure}

The vortex panel method discretizes a body into $\npan$ panels.
Two points on each panel $p$ are of particular interest: (1)~the panel quarter-chord point will 
correspond to the location of a point vortex with unknown strength $\pangam_p$ 
(i.e.,~the point at which the panel's bound vorticity is confined), and 
(2)~the panel three-quarter-chord point will represent the point at 
which the no-flow-through boundary condition will be imposed (i.e.,~the ``collocation point'').
Note that the lumped vortex panel method enforces 
the Kutta condition by construction, since the circulation at the trailing 
edge is always exactly zero.
Further, Kelvin's circulation theorem is enforced through the presence of 
a ``starting vortex'' with strength $(-\sum_{p=1}^{\npan}\pangam_p)$ that resides in the far-field.
In the present development, the only external flow influences to be considered are the 
freestream velocity $\uinf$ and the velocity induced 
by a nearby wake $\uwake$.
Taking $\uinf$ and $\uwake$ to be evaluated at each of the collocation points $(x_c,y_c)$,
we impose the flow tangency condition at each collocation point:
 \begin{equation}
   \sum_{p=1}^\npan a_{cp}\pangam_p  = -\left(\uinf + \uwake\right)\cdot\pannorm_c,
\label{eq:noflowthrough}
 \end{equation}
where $\pannorm_c$ is the outward unit normal vector at collocation point $c$ and
\begin{equation}
a_{cp} = \frac{\pangam_p\left(y_p-y_c,\ x_c-x_p\right)}{2\pi\left((x_c-x_p)^2+(y_c-y_p)^2\right)}
\cdot\pannorm_c
\end{equation}
is an ``influence coefficient'' that represents the normal velocity induced on 
collocation point $c$ due to unit bound circulation on panel $p$.
The panel strengths $\pangam_p$ can be determined by solving the system of $\npan$ 
linear equations in~\eqref{eq:noflowthrough}.
%
%
An example of streamlines generated by the combined wake-body model are shown in Figure~\ref{fig:fishwake}.
Note that we assume the swimmer is coasting here, and so have no included the influence of 
body motions in our formulation; such considerations can be included by appropriately modifying 
the right-hand-side of Eq.~\eqref{eq:noflowthrough}.


We now describe how to determine the velocity $\uwake$ induced by a nearby vortex wake on 
each collocation point.
For each vortex $\alpha$ among the $N$ in the strip, denote the position as 
$z_\alpha=x_\alpha+iy_\alpha$ and the strength as $\Gamma_\alpha$.
Then, the velocity induced by the singly-periodic system of point vortices at 
the collocation point $(x_c,y_c)$ is given by,
\begin{equation}
 \uwake=\sum_{\alpha=1}^N\frac{\Gamma_\alpha\left(-\sinh(\Delta y_\alpha),\ \sin(\Delta x_\alpha)\right)}{2L\left(\cosh(\Delta y_\alpha) - \cos(\Delta x_\alpha) + \epsilon\right)}
\label{eq:velo_total}
\end{equation}
where $\Delta x_\alpha=\twopiL(x_{c}-x_\alpha)$, $\Delta y_\alpha =\twopiL(y_{c}-y_\alpha)$, 
and $\epsilon$ is a regularization parameter that is used to ensure smoothness of solutions.
As the wake evolves according to~\eqref{eq:vort_eom}, the induced velocity $\uwake$ at each collocation point 
will change accordingly, and is thus a function of time.
As such, the panel strengths will also be a function of time and~\eqref{eq:noflowthrough} must be 
solved at each time-step.
In combining the vortex panel method with 
the idealized wake models, we have assumed that the wake induces a velocity on the body, 
but that the body does not alter the wake dynamics.
We employ this simplifying assumption in order to preserve the convenient form of 
the 2S and 2P wake dynamics models described above.

The time-varying nature of the hydrodynamic signature on the fish-like body will be leveraged to detect and classify
wake regimes.
However, the full spatially distributed signature is not often not available for measurement by 
sensors on manmade systems---in fact, as we will see in Section~\ref{sec:results}, such spatially resolved 
measurements may not even be necessary for wake detection and classification in many cases.

According to the recent review in~\cite{triantafyllouAnnRev2016}, most of the research and development of 
bioinspired hydrodynamic reception systems has focused on structures based on the superficial neuromast, which effectively measure local velocities.
Thus, in the remainder of the study, we will focus on time-series measurements of local velocity at a 
single collocation point on the fish-like body.
For the wake detection and classification task, we will work with a signal $\uprime$ that 
corresponds to the tangential velocity at a specified collocation point with 
the tangential component of the freestream velocity subtracted out.
%
%

\section{Results: wake signatures and classification}
\label{sec:results}

With the ideal-flow model established in Section~\ref{sec:modeling},
we are ready to construct a library of wake signatures and assess the efficacy of 
the wake detection and classification protocol.
%

%




\subsection{Wake signatures and measured hydrodynamic signals}
In this study, a NACA~0020 airfoil with chord-length $c$ is used to model a fish-like body.
The airfoil is represented by a total of $\npan=40$ vortex panels.
%
%
%
The dynamic wake-body simulations are performed with a fourth-order Runge-Kutta time-marching scheme;
sensor measurements are sampled uniformly every 0.01 convective time-units.
In all the results presented, the measured hydrodynamic signal $\uprime$ at a collocation point
corresponds to the tangential component of local velocity with the tangential component of the freestream removed.
Unless otherwise stated, the sensor measurements $\uprime$ are taken from the mid-chord on the wake-side of the body.
%
%
%
%

A total of 180 different wake realizations are simulated to build up the library 
and to subsequently assess the performance of the wake detection protocol.
Multiple realizations of wake signatures are collected from each of the 2S and 2P wake regimes 
(i.e.,~vK, rvK, E1, O1, and O2).
For the study at hand, 
we take care to ensure that some consistency is maintained
between the different wakes that are considered.
Wake parameter values are chosen such that all 2S and 2P wakes have vortices with comparable strengths;
all 2S wakes are constructed with strengths in the range $|\Gamma|\in[0.3,0.5]$, while
%
%
all 2P wakes are generated  with $\gamma = 3/7$.
For the 2P wakes, we also take ${\mathcal{P}\in\{0,-0.55,-0.605,-0.71,-0.803,-0.85,-0.9, -0.95\}}$,
such that the typical $E1$, $O1$, and $O2$ regimes are realizable.
%
%
%
%
Further, the lateral separation between the wake and body must be consistent,
in order to draw a fair comparison; otherwise, some signatures may be 
stronger or weaker by virtue of this geometric parameter, 
rather than by the nature of the wake dynamics.
To this end, we also note that the vortex trajectories of 2P wakes lead to a time-varying lateral separation between the wake vortices 
and the body-axis; 
thus, the separation between the wake-axis and the body-axis may not be the most relevant parameter to keep fixed between 
realizations if an objective assessment is desired.
Instead, we define $h$ to be the minimal lateral separation achieved between wake vortices and the body-axis, over the course of the wake evolution.
Then, we take $h=0.2L$ across all wake-body simulations studied here.
Figure~\ref{fig:fishwake} shows a representative snapshot
of the fish-like body in the vicinity of a dynamic 2P wake.
Although the vortices appear to have a substantial lateral separation from the body-axis 
in this snapshot,
the lower row of vortices will reach a minimal separation $h=0.2L$ with the body-axis 
at some point in each cycle of their motion.
%
%

\begin{figure}[h!]
\begin{center}
\subfloat[vK]{
 \includegraphics[width=0.3\textwidth]{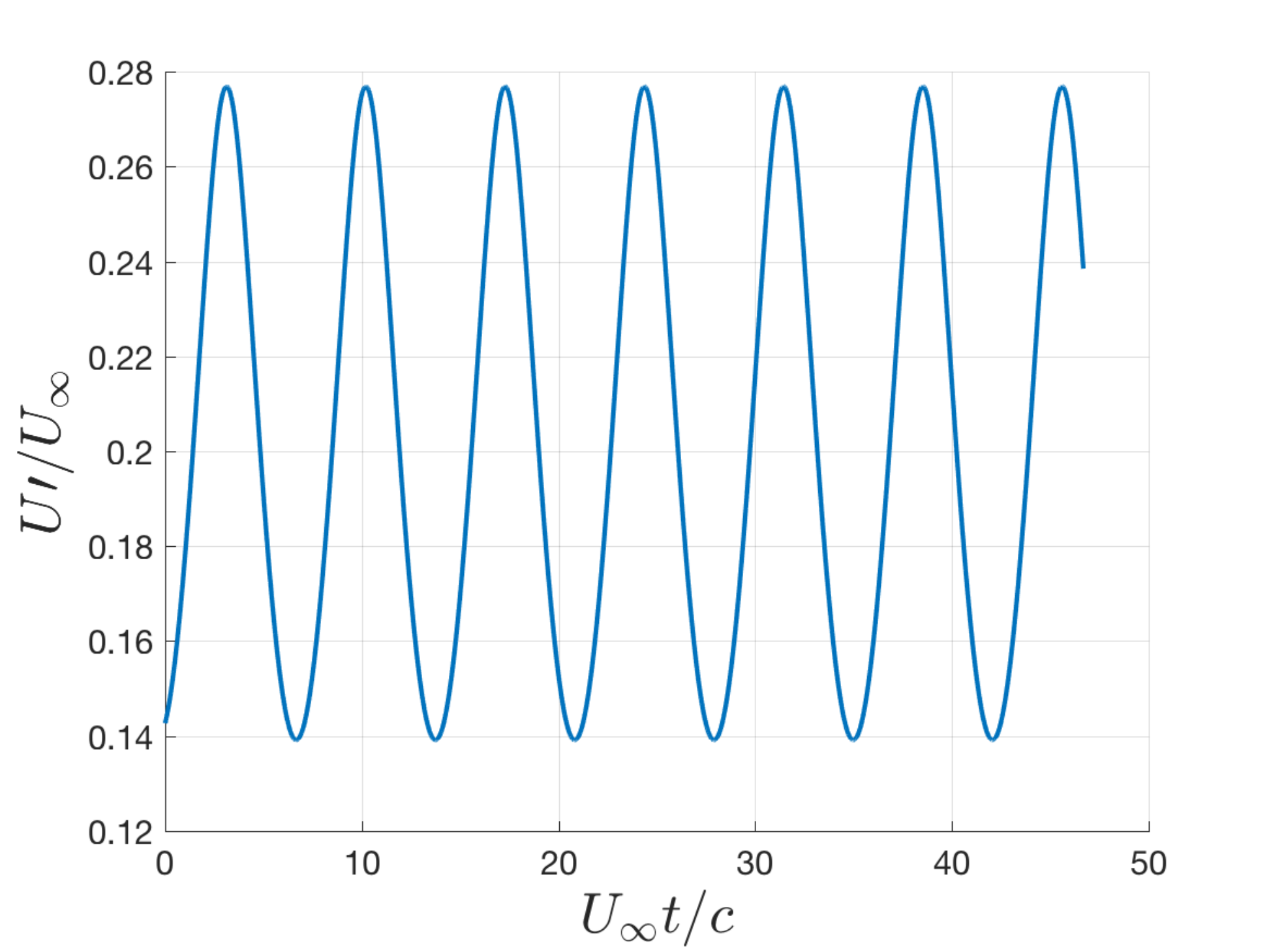}
} \label{fig:velowake2Svk}
\subfloat[rvK]{
 \includegraphics[width=0.3\textwidth]{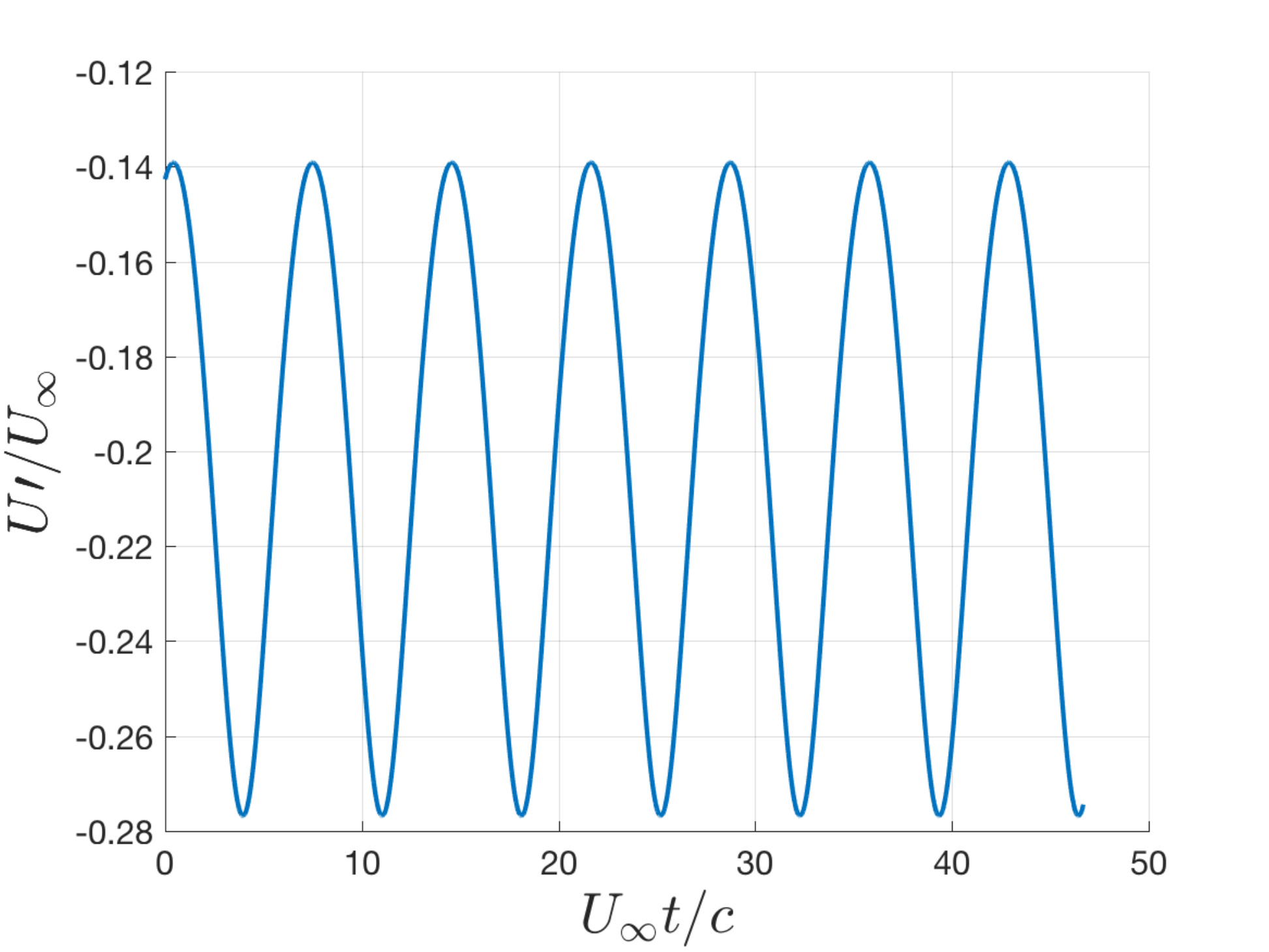}
} \label{fig:velowake2Srvk}\\
  \subfloat [E1]{
 \includegraphics[width=0.3\textwidth]{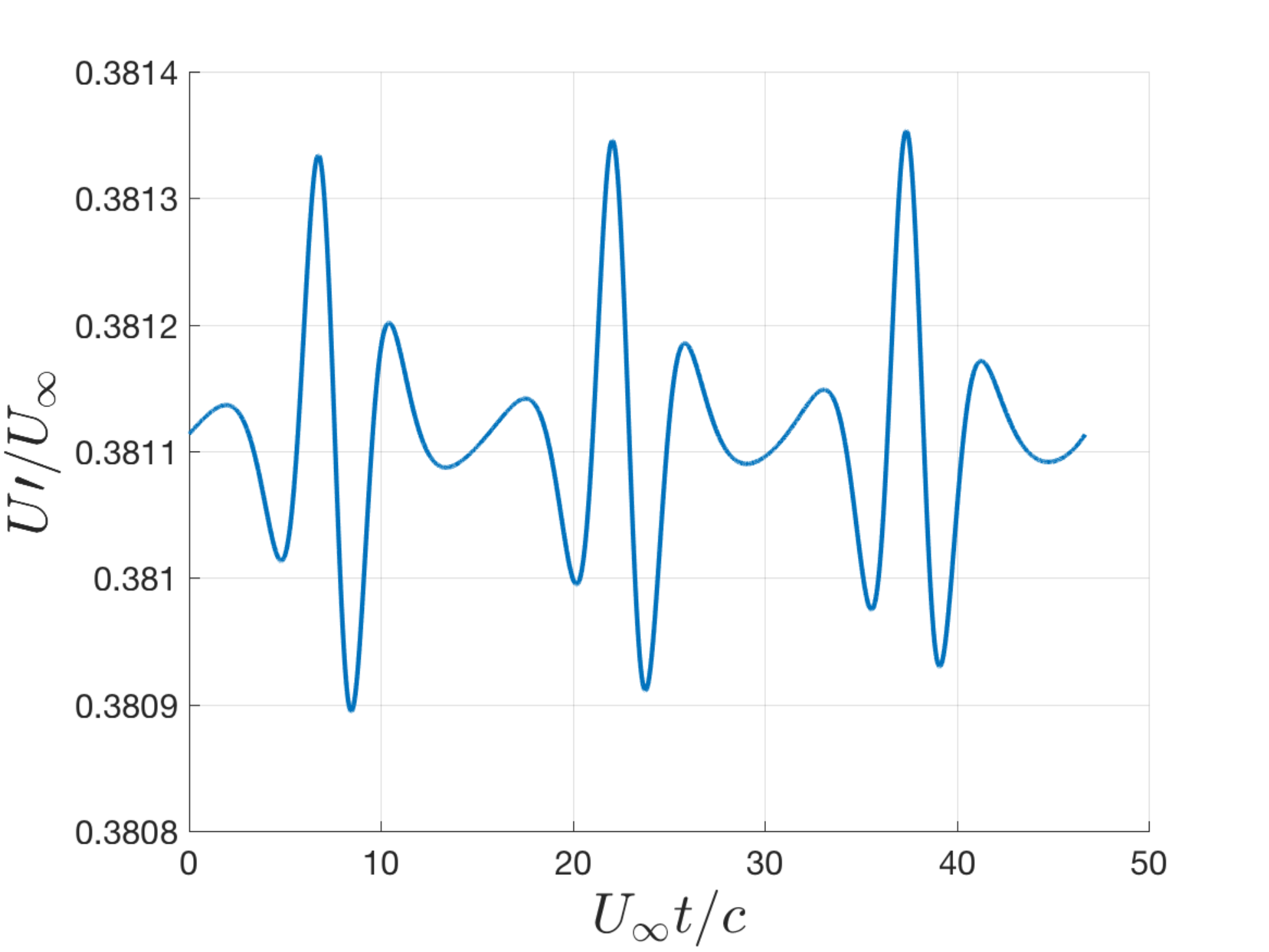}
 }  \label{fig:velowakeE1}
\subfloat[O1 ]{\includegraphics[width=0.3\textwidth]{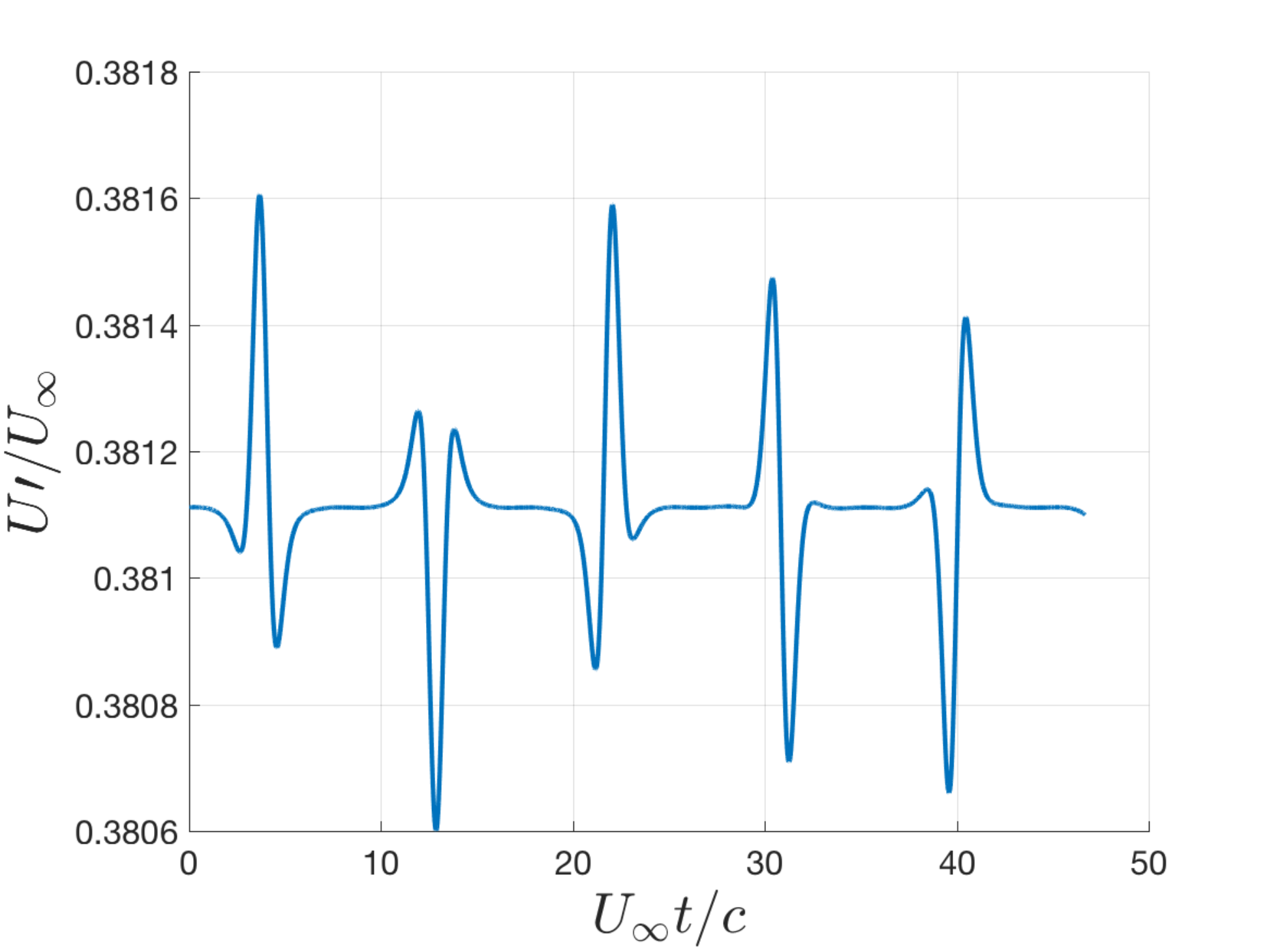}
} \label{fig:velowakeO1}
 \subfloat [O2]{
 \includegraphics[width=0.3\textwidth]{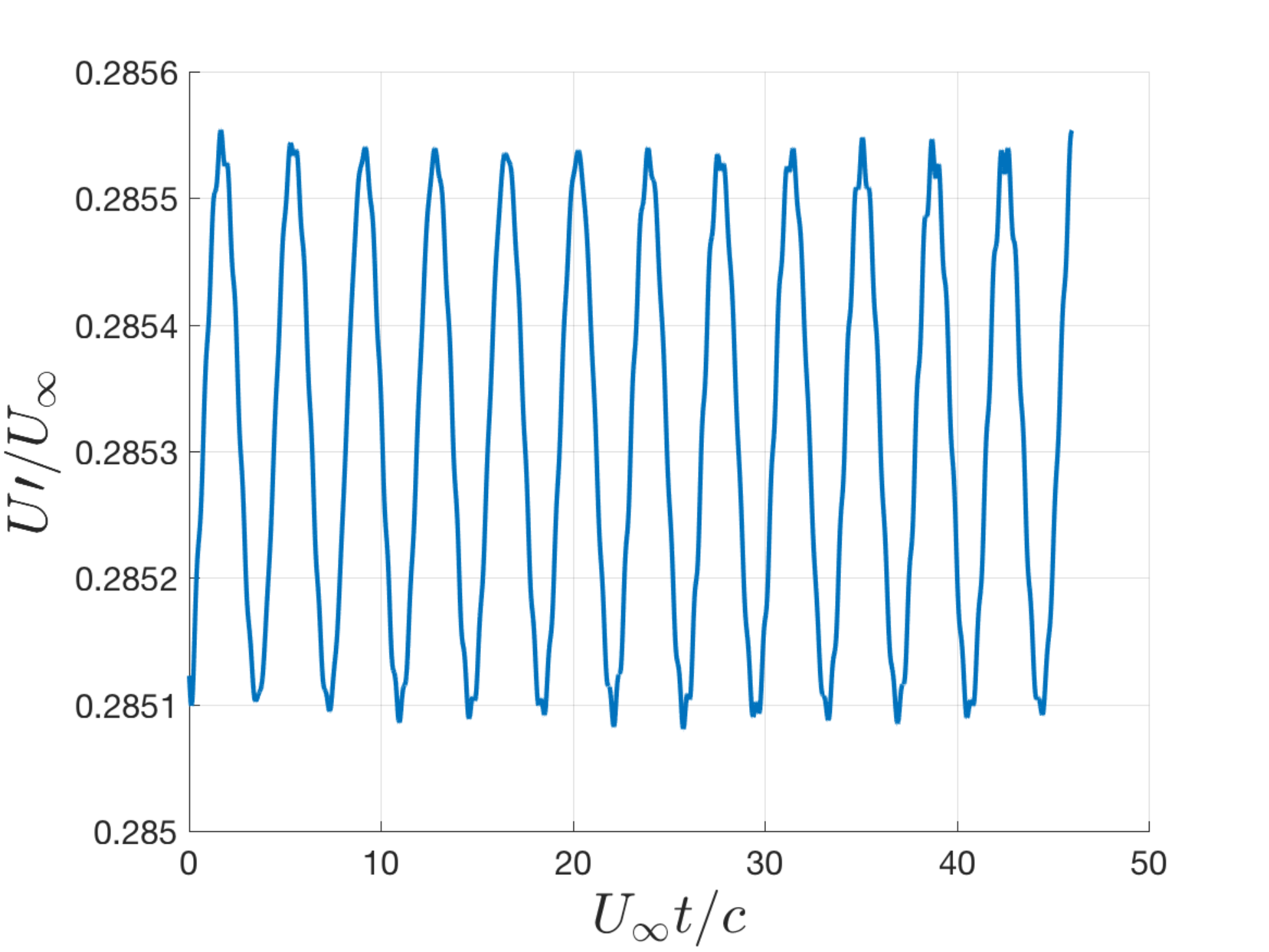}
 }   \label{fig:velowakeO2}
 \caption{Sensor signals $\uprime$ have notable qualitative differences between different wake regimes. The representative signals shown here correspond to (a) $\Gamma = 0.4$ (b) $\Gamma = -0.4$ and (c)--(e) $\gamma = 3/7$, $\mathcal{P} = -0.803$ with $\mathcal{H} =-0.2, 0.4,$ and $0.13$, respectively.} \label{fig:velowake}
 \end{center}
\end{figure}

Representative time-series measurements $\uprime$ for the different wake regimes considered are shown in Figure~\ref{fig:velowake}. 
With the exception of vK and rvK wakes, the signals appear to be qualitatively different from one another;
indeed, the signals appear to have different frequency content.
This suggests that a feature vector based on a frequency-domain representation of the signal
may be well-suited for the classification task.
In the case of vK and rvK, the sign of $\uprime$ can be used to distinguish between these wakes, 
as will be discussed later.
%
%
In the next section, we describe a specific feature extraction method that can be used to build up 
an effective wake classification library.
%

\subsection{Feature extraction from frequency-domain signatures}

\begin{figure} [h!]
	\begin{center}
    \subfloat [vK]{
    \includegraphics [width=0.3\textwidth]{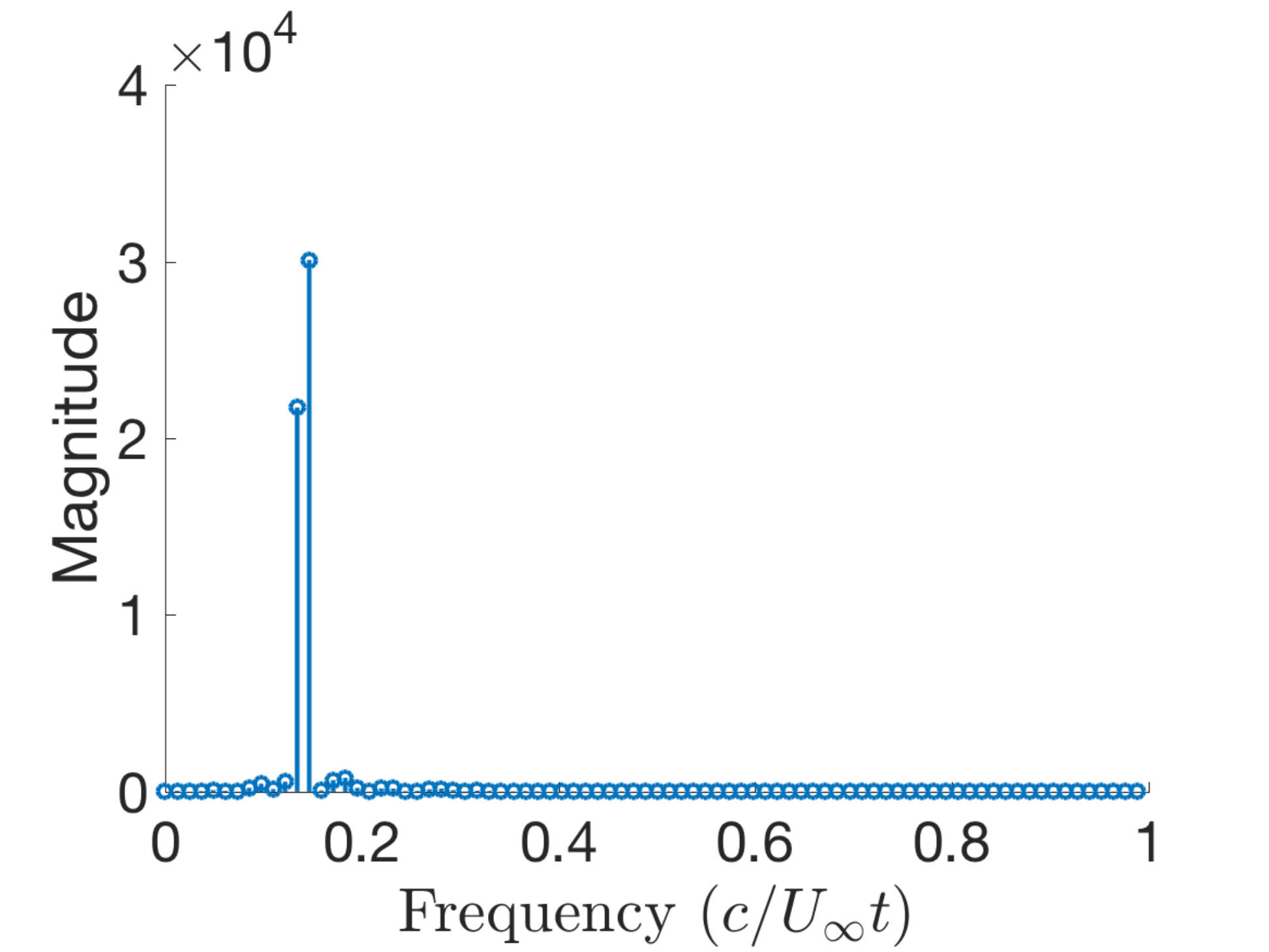}
 }  \label{fig:FFTvK}
 \subfloat [rvK]{
    \includegraphics [width=0.3\textwidth]{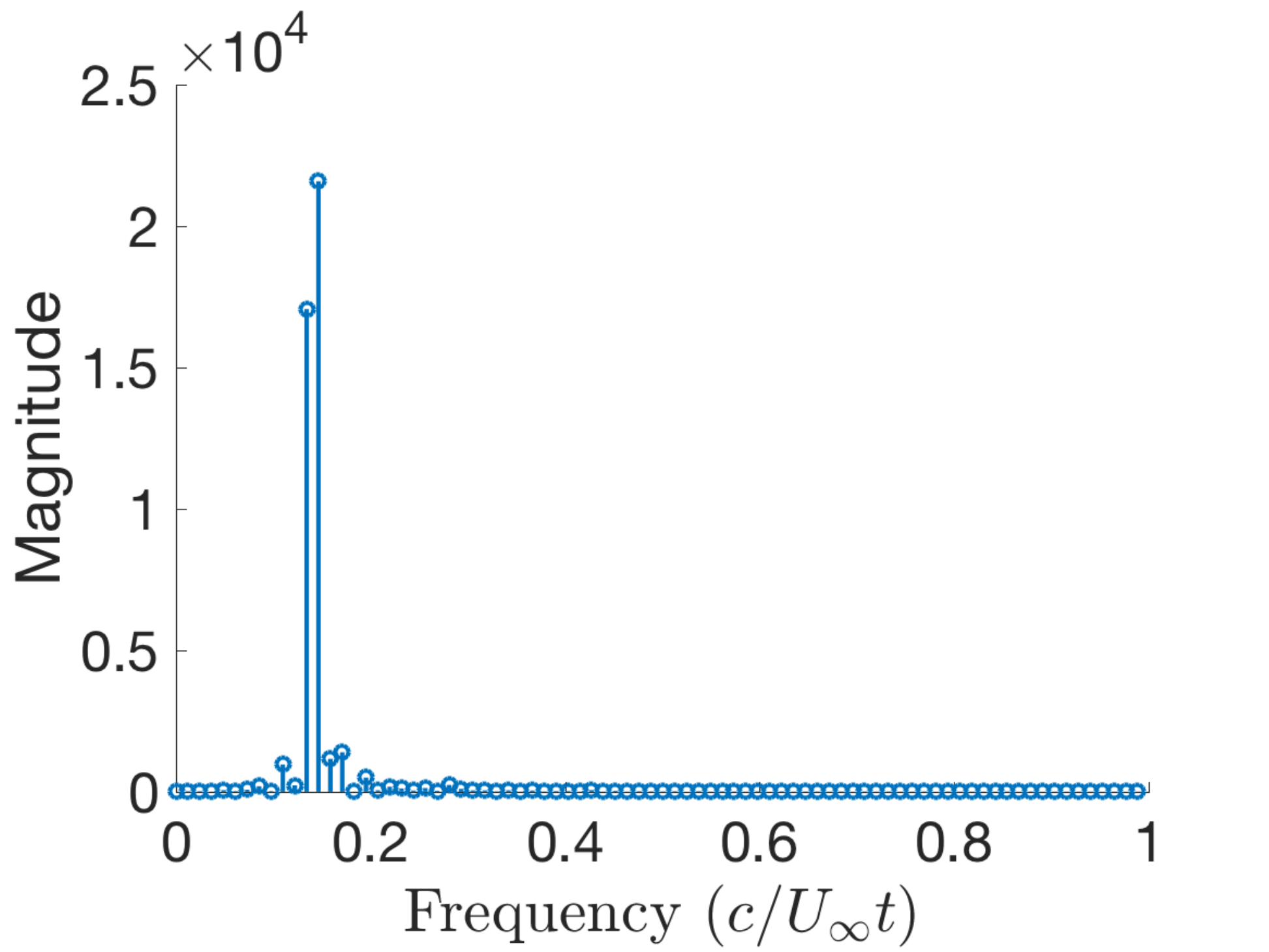}
 }  \label{fig:FFTrvK}\\
    \subfloat [$E1$] {
    \includegraphics [width=0.3\textwidth]{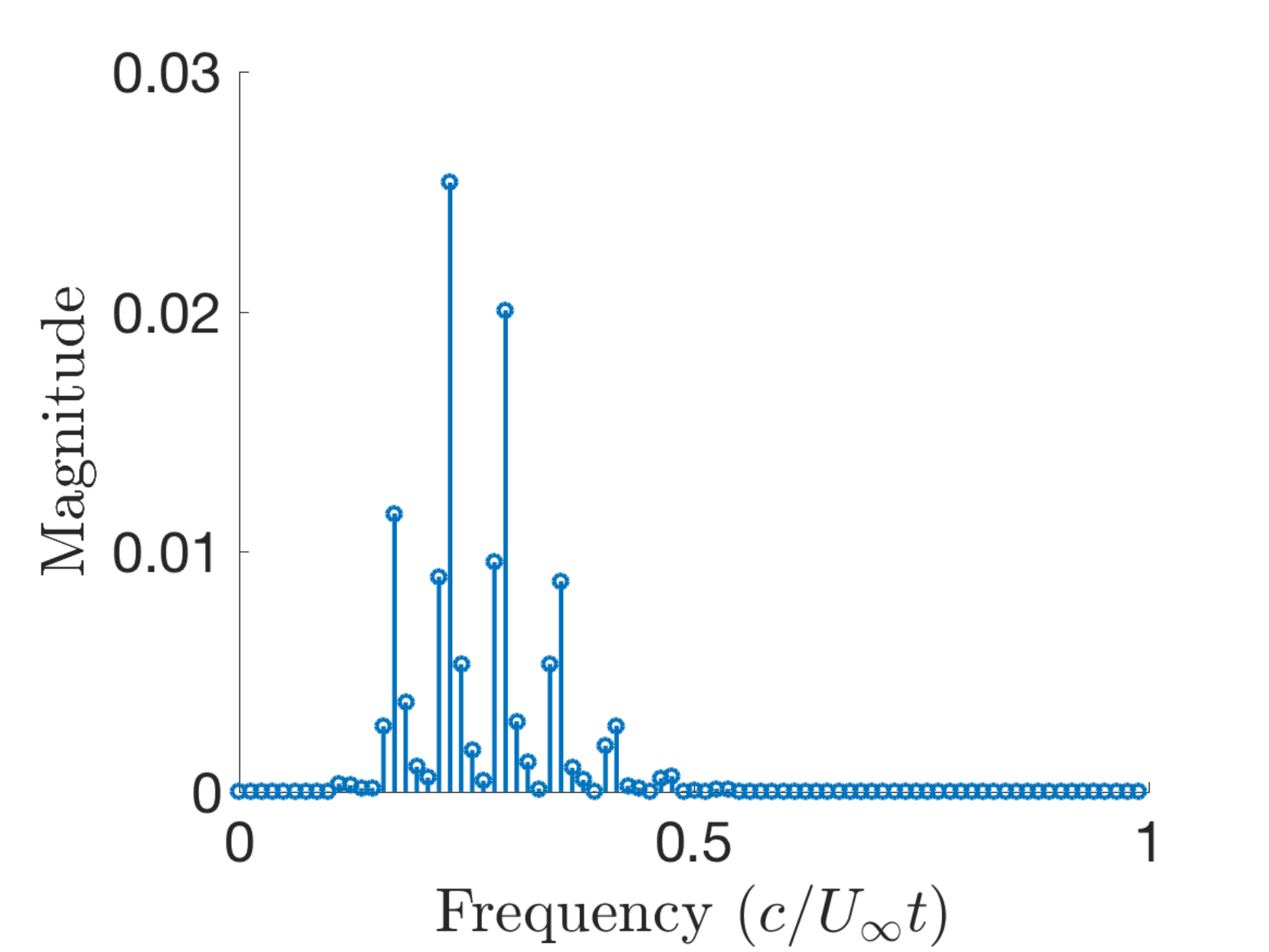}
 }  \label{fig:FFTE1}
    \subfloat [$O1$]{
    \includegraphics [width=0.3\textwidth]{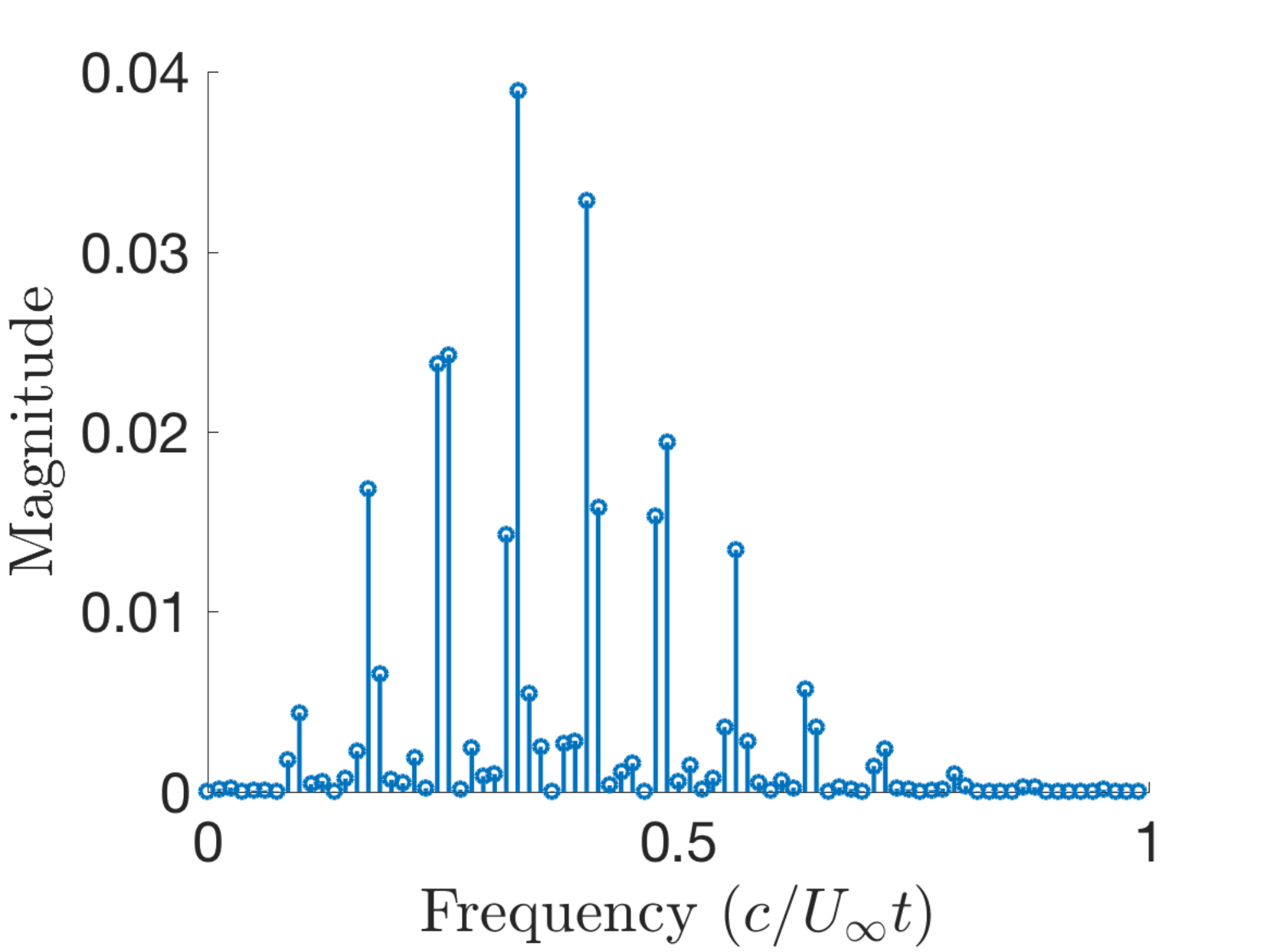}
 }  \label{fig:FFTO1}
    \subfloat [$O2$] {
    \includegraphics [width=0.3\textwidth]{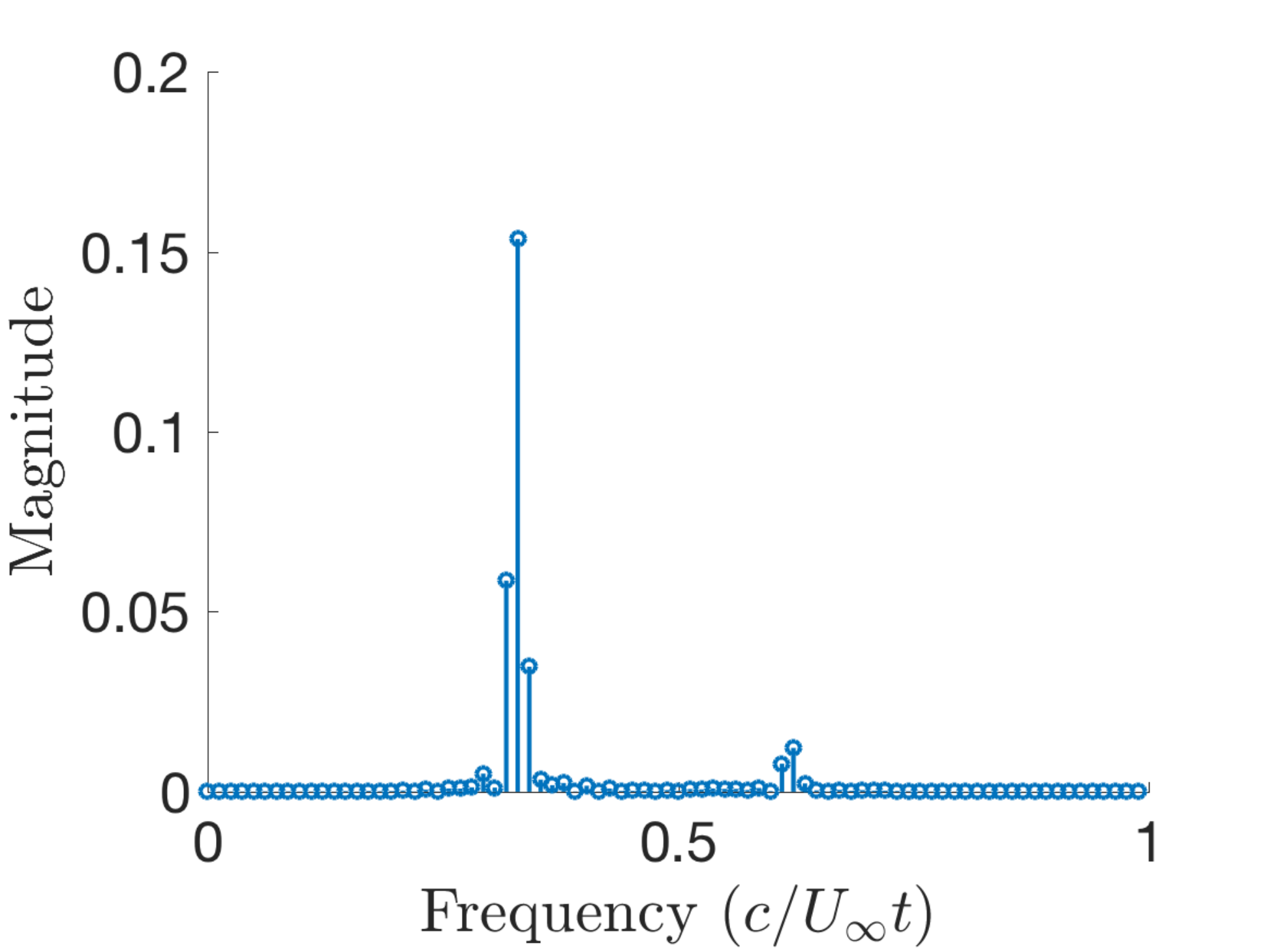}
 }   \label{fig:FFTO2}
\caption[]{Different wake regimes impart distinct signatures on the fish-like body, which can be seen in the frequency-domain. The frequency-domain signatures shown here correspond to spectra from the corresponding plots in~Figure~\ref{fig:velowake}.}
\label{fig:FFT}
\end{center}
\end{figure}

As seen in Figure~\ref{fig:velowake}, different wake regimes impart distinct hydrodynamic signatures on the body, which can be characterized by differences in frequency content.
%
To verify this, we compute the frequency spectra of each signal in Figure~\ref{fig:velowake}
by means of a 
fast Fourier transform~(FFT).
Note that frequency content can also be efficiently computed on-line when the sensor measurements are in the form 
of a datastream, as is needed in practical real-time applications.
The frequency-domain representations of each of these representative signals, shown in Figure~\ref{fig:FFT},
show qualitative differences between each wake regime as well---again with the exception of 
the vK and rvK wakes.
For example, 2S wakes exhibit a single dominant frequency with a large amplitude,
whereas 2P wakes are more broadbanded;
this feature alone can be enough to distinguish a 2S wake from a 2P wake.

\begin{figure} [h!]
\begin{center}
    \includegraphics [width=0.6\textwidth]{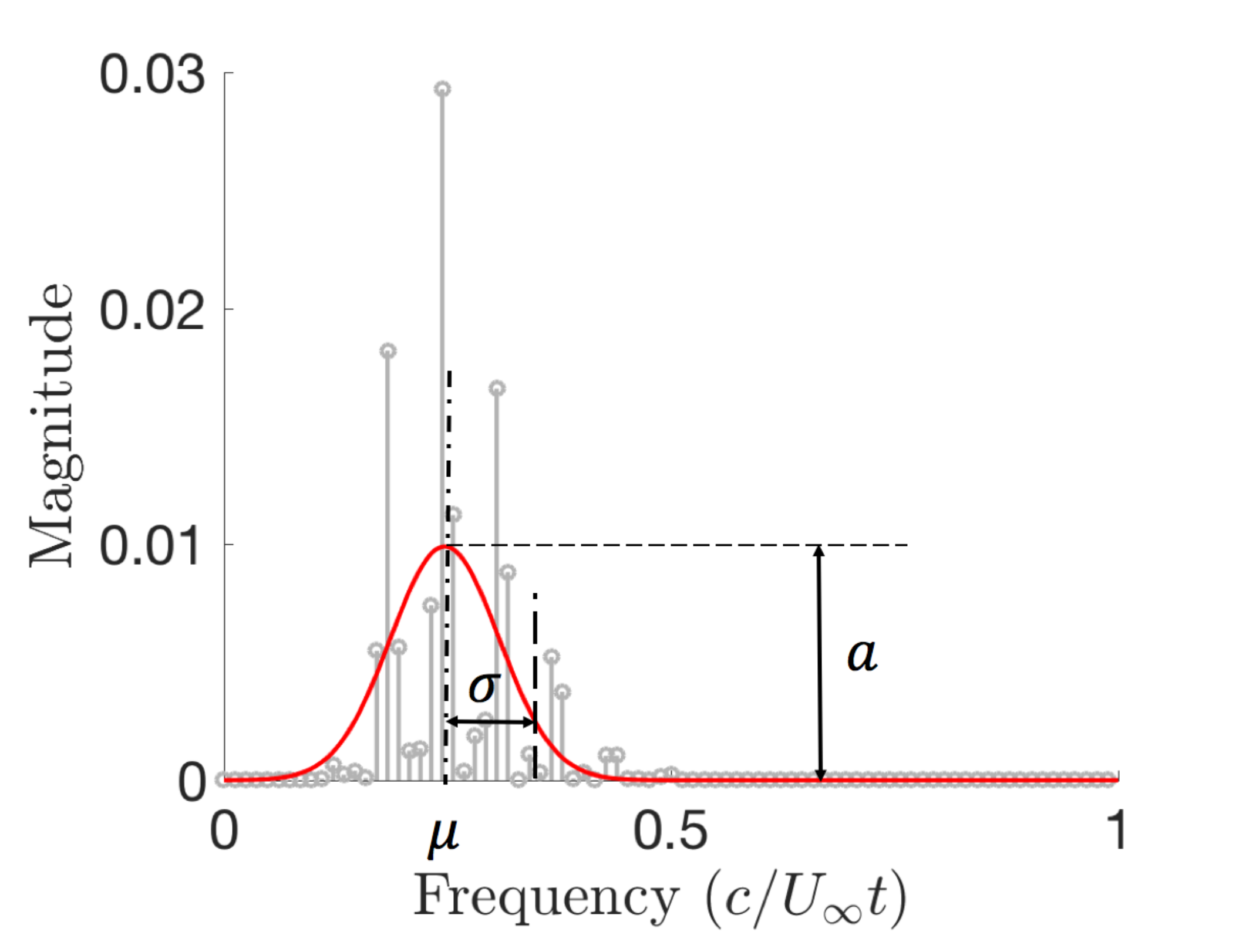}
\end{center}
\caption{A Gaussian fit to the frequency signature yields a convenient, concise, and effective set of parameters $(\mu,a,\sigma)$ for use as a feature vector in the wake classification task~(see Eq.~\eqref{eq:bell}).  The example plotted here corresponds to the $E1$ signature in tile (c) of Figure~\ref{fig:FFT}.  The parameters of the best fit Gaussian bell curve (drawn in red) concisely summarize this signature as a feature vector.}
\label{fig:GFex}
\end{figure}

The frequency-domain representation is a convenient one compared to the time-domain representation of the signal 
for wake detection because---all else equal---it is less sensitive to implementation details, 
such as measurement start time, sampling rate, and sampling duration.
Thus, we will extract feature vectors from each signal's frequency-domain representation.
The frequency signatures in Figure~\ref{fig:FFT} each display a distinct ``bell'' shape.
Thus, we choose to summarize the frequency-domain signatures by fitting a Gaussian bell curve,
\begin{equation}
F(x) = a\exp{\left(-\frac{ \left( {x - \mu } \right)^2 } {2\sigma ^2 }\right)}.
\label{eq:bell}
\end{equation}
Note that the specific shape of Gaussian bell curve is fully defined by the parameters $(\mu,a,\sigma)$, 
which are determined by the fitting procedure.
An example is shown in Figure~\ref{fig:GFex}.
A feature vector for each wake realization $i$ can then be defined using the parameters of the associated best-fit Gaussian bell curve: 
\begin{equation}
V_i=(\mu_i,a_i,\sigma_i).
\label{eq:featvec}
\end{equation}

\begin{figure}[h!]
 \centering
 \includegraphics[width=\textwidth] 
 {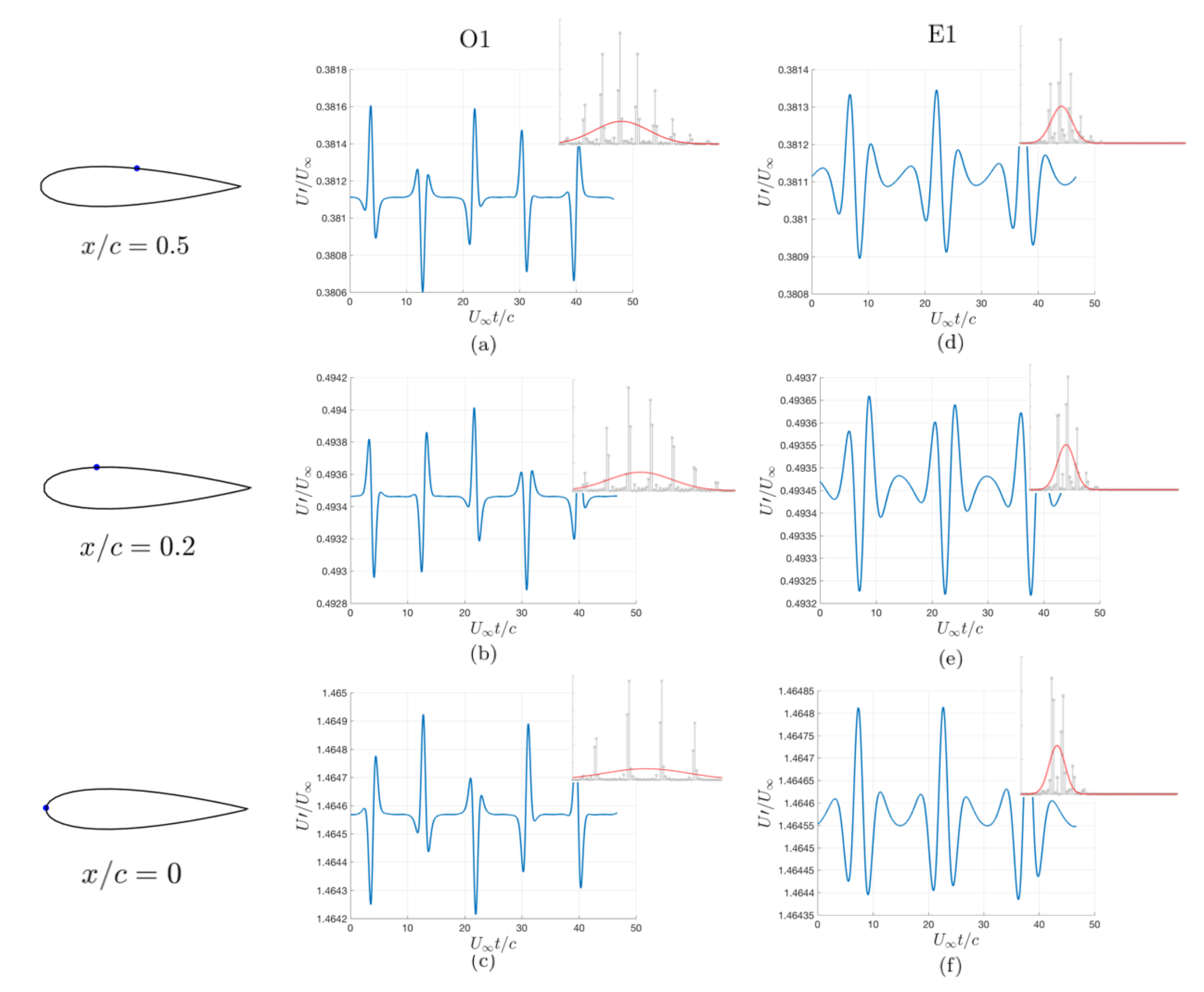}
\caption{Wake signatures measured at different locations along the fish body are qualitatively similar. Time- and frequency-domain signatures and corresponding feature vectors are presented for $E1$ and $O1$ wakes at the three sensor locations shown.  The $E1$ and $O1$ wake signals are generated based on the same wake parameters used in Figures~\ref{fig:velowake} and \ref{fig:FFT}.}
\label{fig:DifferLoc}

\end{figure}

In the next section, we use the feature vector defined in~\eqref{eq:featvec} to build up a classification 
library for wake detection,
but first it is worth investigating the influence of sensor location 
on the measured signals and the resulting feature vectors.
One might suspect that sensor signals will vary significantly from one measurement location to another, especially for the more complicated wake signatures of regimes $E1$ and $O1$; 
however, a qualitative comparison of the time-series signals from different sensor locations reveals only slight differences.
Figure~\ref{fig:DifferLoc} compares hydrodynamic signals imparted by $E1$ and $O1$ 
wake regimes measured at the mid-chord ($x/c=0.5$), the maximum thickness point ($x/c=0.2$), and the leading edge ($x/c=0$).
The frequency-domain signatures and associated feature vectors are included as insets to Figure~\ref{fig:DifferLoc}.
Although the feature vectors will have quantitative differences between them, 
the qualitative similarity between feature vectors indicates that our proposed 
feature extraction approach is an appropriate 
one regardless of sensor location.
For brevity, in the remainder we only present results for the mid-chord sensor location $x/c=0.5$.

\subsection{Wake Classification}
At this point, a classification library can be built up using the feature vectors defined in~\eqref{eq:featvec}.
The library is constructed by collecting an equal number of distinct feature vectors from each wake regime.
The classification library with 130 entries is presented in Figure~\ref{fig:KNNlib}.
We will report on the influence of ``training size'' (i.e.,~the number of library entries) on classification outcomes momentarily.

\begin{figure} [h!]
 \begin{minipage}{0.6\textwidth}
	\centering
	\subfloat[Full Library]{
    	\includegraphics[width=\textwidth] {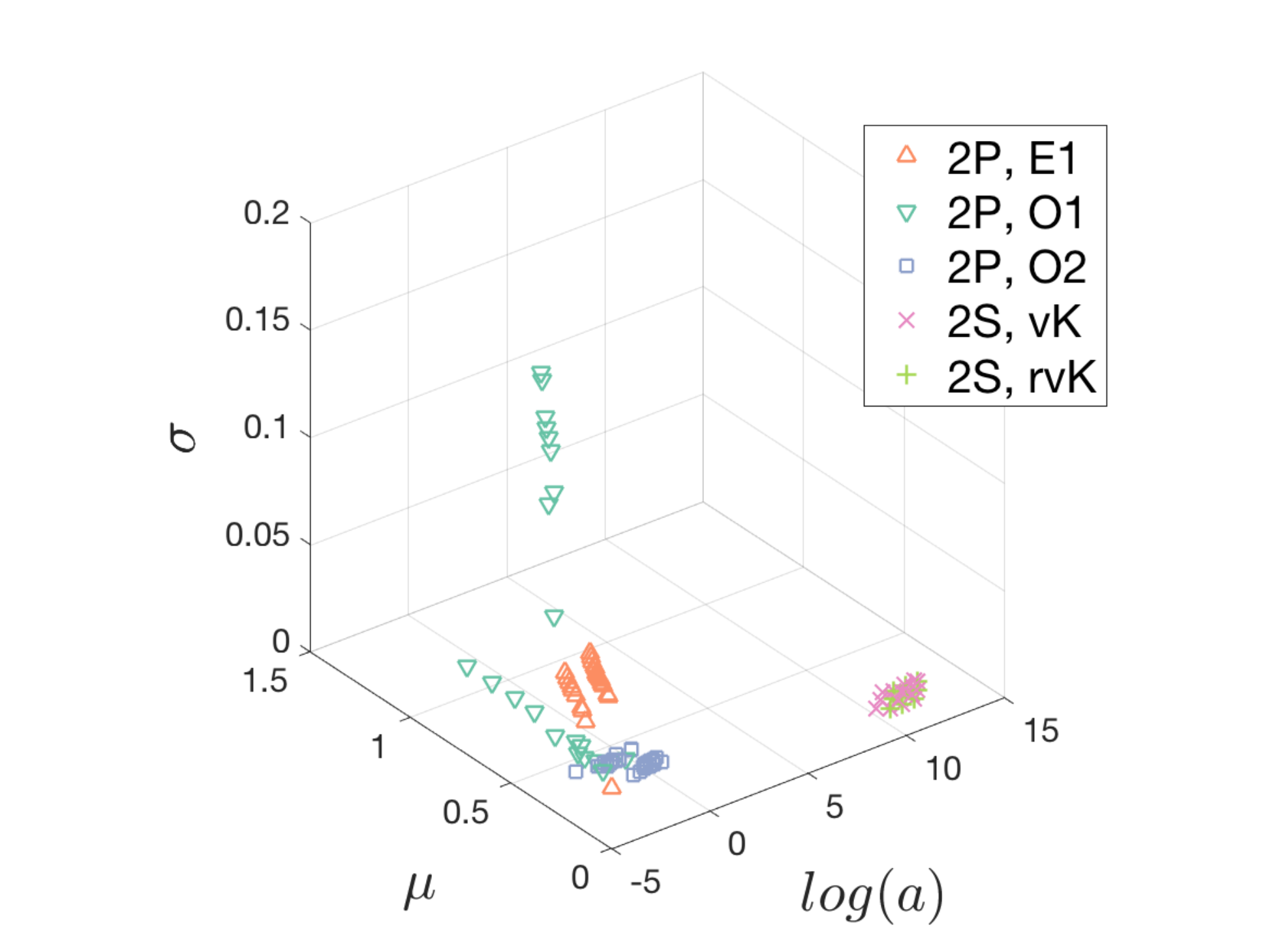}
        \label{fig:KNNlib}
    }
   \end{minipage}\hfill
   \begin{minipage}{0.4\textwidth}
   \centering
   \subfloat[2P cluster in the library]{
   		\includegraphics[width=\textwidth]{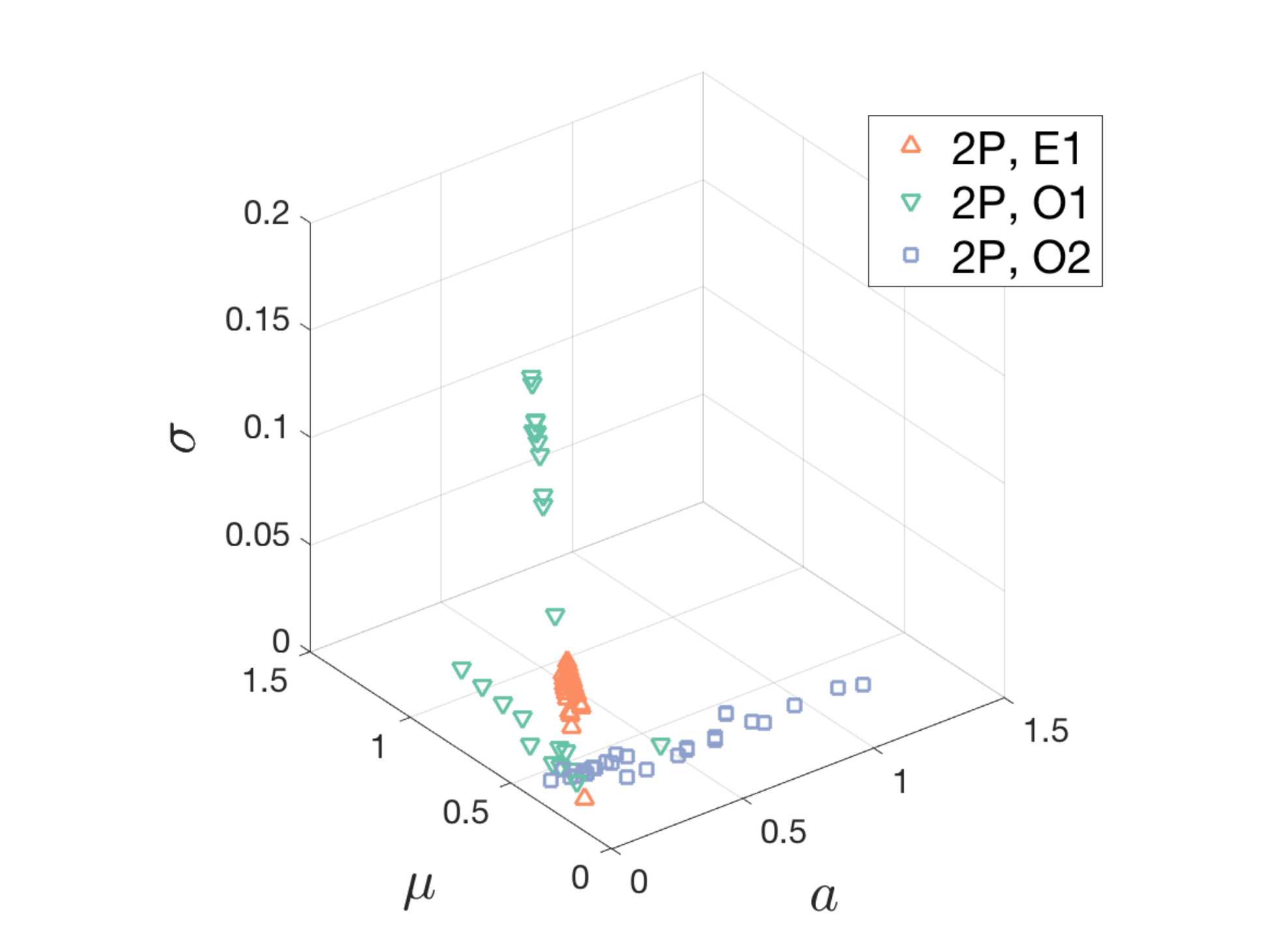}
        \label{figure:2Plib}
   }\hfill
   \subfloat[2S cluster in the library.]{
   		\includegraphics[width=\textwidth]{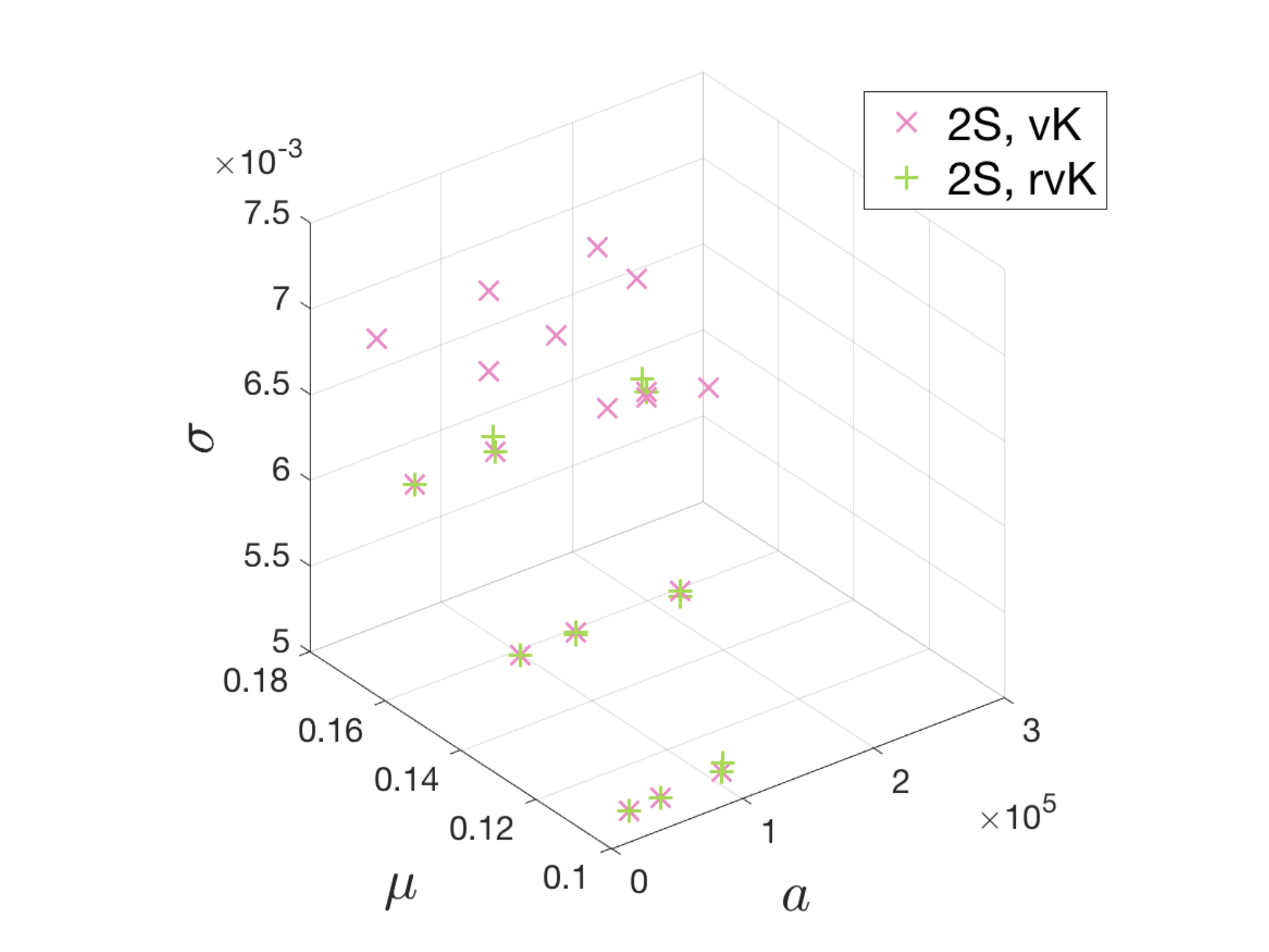}
        \label{figure:2Slib}
  }
   \end{minipage}
\caption{The wake classification library is built up in feature space with library entries from many wake types and regimes. In (a), an example of a classification library with a total of 130 entries is presented, equally distributed between the five regimes considered.  The 2S and 2P wakes separate into distinct clusters here.  A blown-up plot of the 2P and 2S clusters are shown in (b) and (c), respectively.  Note that the 2P regimes are fairly well separated, whereas the 2S regimes are blended together.}
\label{fig:KNNlib}
\end{figure}

A quick study of Figure~\ref{fig:KNNlib} reveals that 2S and 2P wake types are fully partitioned into distinct clusters in feature vector space.
A closer look at the 2P clustering in Figure~\ref{figure:2Plib}
shows that the various 2P wake regimes also partition relatively well, 
though some overlap between wake regimes is observed.
Interestingly, many of these overlapping points correspond to wake signatures generated close to regime boundaries.
In contrast to the 2P wake regimes, the partitioning of the vK and rvK configurations of 2S wakes do not partition 
as nicely (see Figure~\ref{figure:2Slib})---as is expected based on our previous 
observations of the associated signatures in both the time- and frequency-domain.

As mentioned in the development of the general wake detection framework of Section~\ref{sec:wakedetection},
once a library of feature vectors is available, numerous alternative classification algorithms can be applied
to determine the likelihood that the signature from an unknown wake belongs to a particular class of wakes stored in the library.
Here, we make use of the relatively simple $k$-nearest-neighbor~(KNN) algorithm~\cite{bishop2006,tan2006}.
Given a user-specified value $k$, the KNN algorithm uses a distance measure to determine which $k$ entries among 
the set of all library entries $\{V_1,V_2,...,V_r\}$ are closest to the test vector $V_\text{test}$.
Once the $k$ nearest neighbors have been determined, a majority voting procedure is used 
to assign a label to the test vector;
that is, the unknown wake is determined to belong to the same class as the most frequent class 
among the set of $k$ nearest neighbors.
In the present study, we use the Euclidean distance as a measure of distance for 
the KNN algorithm; 
this assumes that each element of a feature vector has an equal contribution
to characterizing a wake regime.
%

%

\begin{figure} [h!]
	\begin{center}
    \includegraphics [width=0.6\textwidth]{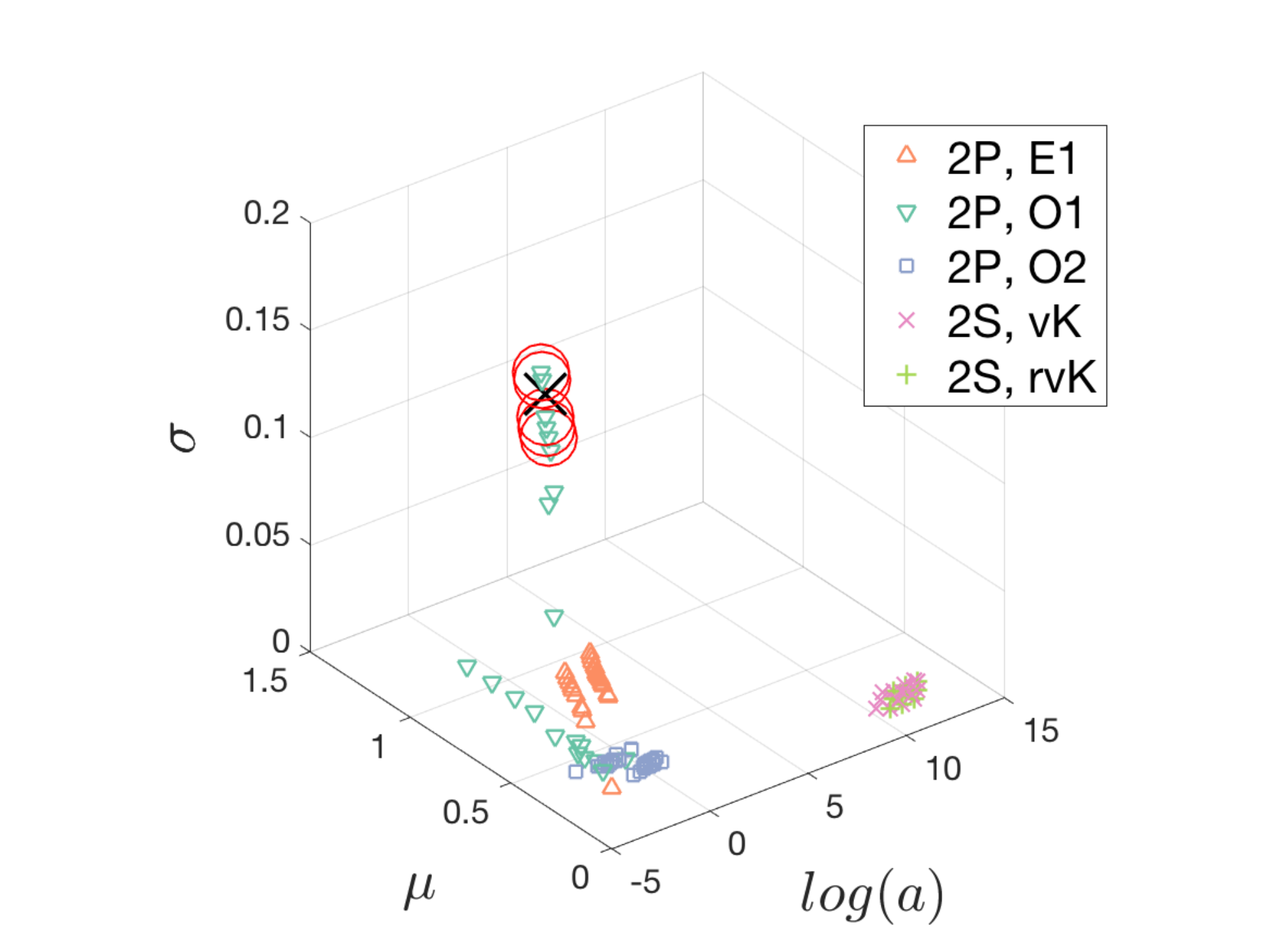}
\caption{An example of the wake library from Figure~\ref{fig:KNNlib} used for classification with a KNN algorithm applied to a test signal (denoted by $\times$).  The $k=5$ nearest neighbors are circled.  The test signal is correctly classified as $O1$ in this example.}
\label{fig:SampleTesting}
\end{center}
\end{figure}

We show a simple example of the KNN classification procedure at play in Figure~\ref{fig:SampleTesting}.
Here, the test vector is denoted with a black $\times$, and the  $k=5$ nearest neighbors 
are circled in red. 
For the example shown, all $k=5$ nearest neighbors correspond to $O1$ wakes, and thus the measured hydrodynamic signal is determined to also be an $O1$ wake.

\begin{figure}[h!]
\begin{center}
  \includegraphics [width=0.7\textwidth] {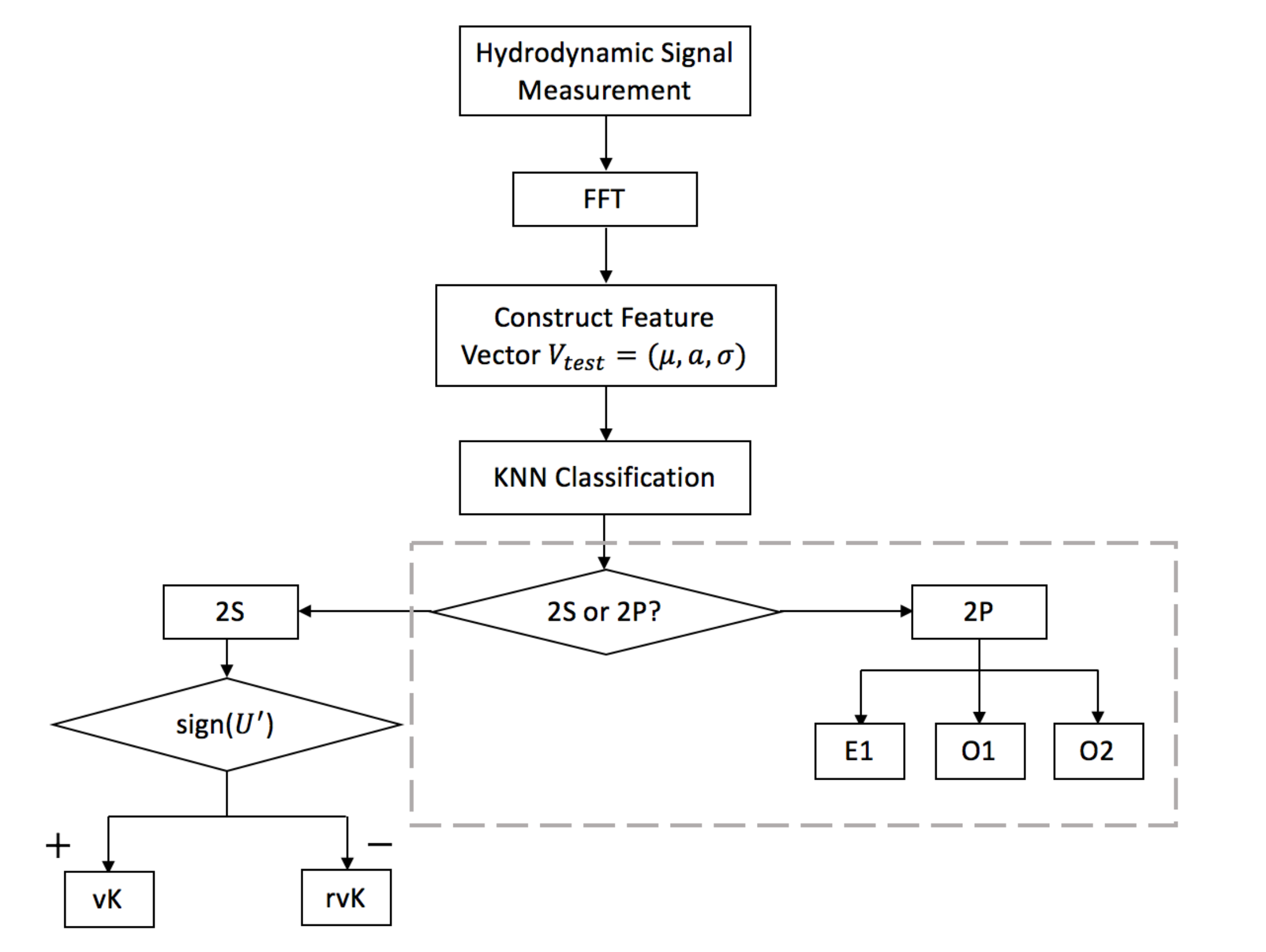}
\end{center}
\caption{A summary of the wake detection protocol used in the proof-of-concept portion of this study.}
\label{fig:frame}
\end{figure}

We now return to the issue of ``blended'' clusters of vK and rvK feature vectors.
Based on the feature vectors presented in the library, it appears that vK and rvK will be 
virtually indistinguishable from one another.
However, noting the strong separation between 2S and 2P type wakes (note the logarithmic scale of $a$), 
we can say with reasonable confidence that a 2S wake can be distinguished from a 2P wake. 
Thus, once a wake is classified as 2S, we can invoke an additional criterion
to determine whether that particular 2S wake is vK or rvK.
Since the primary distinction between vK and rvK wakes is the 
configuration of oppositely signed vortices,
we can expect that the sign of the time-domain measurements $\uprime$ will be opposite 
between vK and rvK wakes.
Indeed, this happens to be the case in general (see Figure~\ref{fig:velowake} for one example), 
and so if a wake is determined to be of the 2S type by the KNN classification procedure,
an additional step is required to check the sign of $\uprime$ to 
determine whether the wake is vK or rvK.
The full wake detection and classification scheme is summarized in Figure~\ref{fig:frame}.

\begin{figure} [h!]
\begin{center}
      \includegraphics [width=0.75\textwidth] {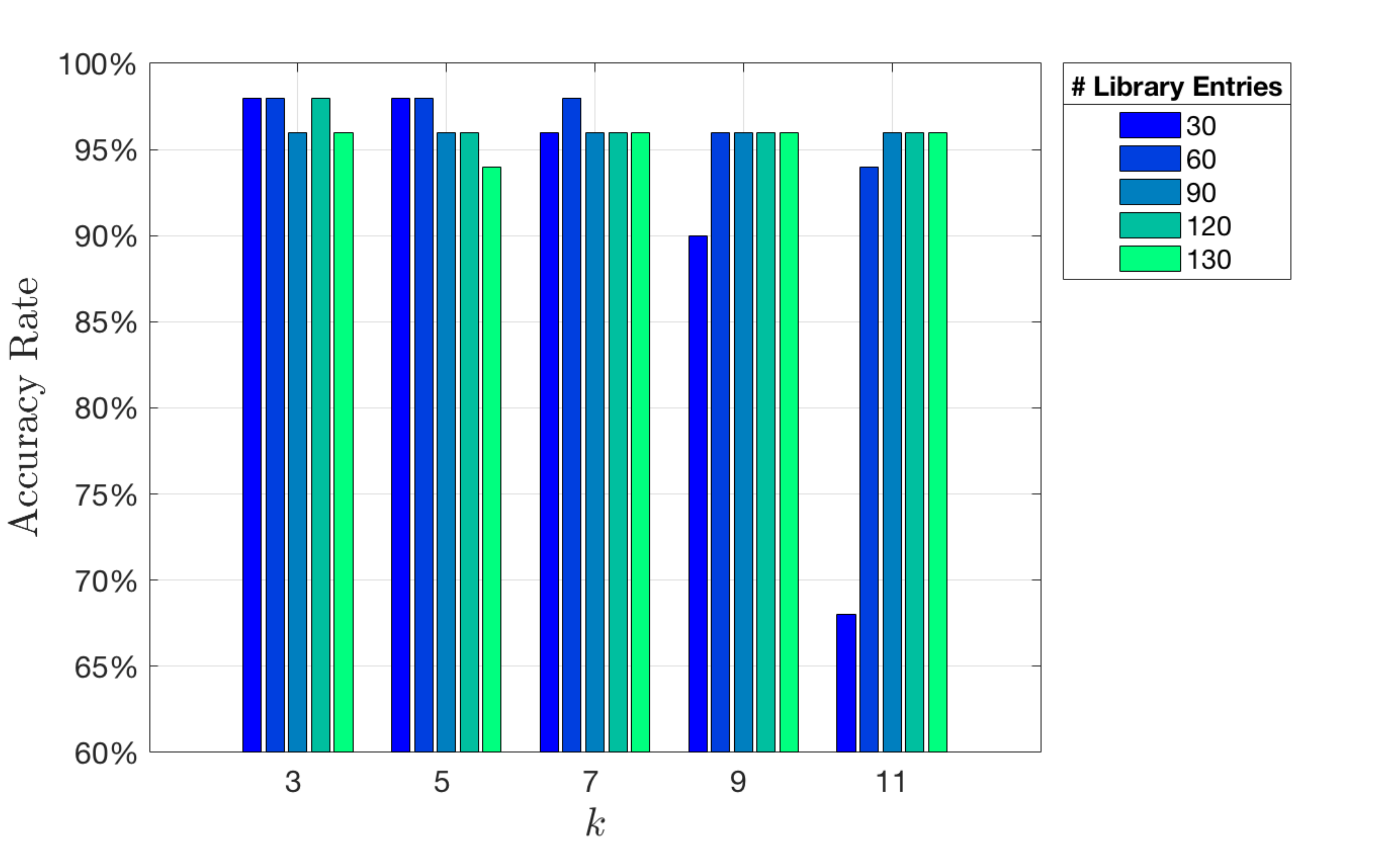}
\end{center}
\caption{A study of accuracy rate as a function of algorithm parameters reveals that the wake detection protocol has good performance.  The number of nearest neighbors $k$ and the number of library entries are both varied in this study.  Accuracy rate is measured based on 50 test signals from randomly generated wakes, and is defined as the number of correct classifications over the total number of test signals.}
\label{fig:KnnAR50}
\end{figure}

In the remainder of this section, we aim to assess the performance of the 
wake detection and classification approach described above.
We are primarily interested in the classification accuracy rate, 
which can be quantified by applying the method to a number of ``test signals''
with known wake regimes.
Then, the accuracy rate can be defined as the fraction of correctly classified test signals
over the total number of test signals.
However, since the accuracy rate will be influenced by various parameters, 
such as the number of nearest neighbors $k$ and the number of entries 
in the classification library,
robustness of accuracy rate to these parameters is the better measure of performance.
Figure~\ref{fig:KnnAR50} shows the accuracy rates corresponding to 
these different parameter values.
In this study, a total of 50 test signals were realized by drawing 
from a random set of 2S and 2P wake models, 
each constructed in a manner consistent with the generation of library entries 
(i.e.,~adhering to the same parameter ranges and constraints).
%
%
The accuracy rate of the wake detection method for the 50 test signals considered 
varies from 68\% to 98\%, depending on $k$ and the number of library entries.
The worst accuracy rate of 68\% corresponds to the case with the smallest number of library entries 
and the largest number of nearest neighbors, as is to be expected.
For all other testing parameters, the accuracy is consistently above 90\%, with accuracy rates above 95\% in the majority of runs.
The high accuracy rate over a broad set of parameters indicates a well-performing 
wake detection protocol.
The strong performance here can be attributed to a suitable feature extraction method,
which was informed by a close examination of representative wake signatures during the
data collection and library construction process.
%

%


\section{Discussion and conclusions}
\label{sec:conclusions}

The results of the present study serve as a proof-of-concept demonstration for the viability of 
exotic wake detection based on hydrodynamic signals.
A dynamic wake model was combined with a vortex panel method to study the hydrodynamic
signatures imparted by various dynamically distinct wake regimes on a fish-like body.
We found that different wake regimes impart distinct hydrodynamic signatures whose qualitative differences
can be detected in time-series measurements taken at a single point on a fish-like body.
Further, we found that the ability to detect these qualitative differences was insensitive 
to the sensor location.

An examination of wake signatures in the frequency-domain revealed that 
a Gaussian bell curve---defined by the parameters $(\mu,a,\sigma)$---could be fit 
to the frequency-signature to provide a concise summary of the wake as a feature vector.
Subsequently, feature vectors from a wide range of wake regimes were collected and
stored for use in a wake classification library.
After doing so, we invoked the $k$-nearest-neighbor algorithm to compare 
the signatures from unknown wake regimes with entries in the wake classification library. 
The performance of this wake detection protocol was assessed by considering 
the accuracy over a range of algorithmic parameters;
this assessment study showed that the wake detection protocol performed with an 
accuracy rate of over 95\% for a majority of algorithm parameter values.
The notable performance of the wake detection method over a wide range of algorithm parameters
indicates that the feature extraction approach was appropriate for summarizing the wake 
signatures;
indeed, the ability of the wake detection protocol to reliably distinguish between wake regimes 
shows that the feature vectors effectively captured the primary 
qualitative distinctions between signatures from dynamically distinct wakes.

One challenge we observed was an inability to reliably distinguish between wake signatures
with wake dynamics evolving near regime boundaries.
A closer inspection of the ``failure cases'' in the performance assessment revealed 
difficulties in distinguishing between wake regimes that evolved close to the 
separatrices in phase-space.
Indeed, these were the same wake regimes that had similar feature vectors 
to one another and were more closely clustered together in the classification library.
This observation suggests that even with a richly populated library, 
the wake detection accuracy rate is unlikely to reach 100\%.
Further, this suggests that other wake regimes not studied here---such as the mixed regime 
reported in~\cite{basuPOF2015} that exhibit qualities of both orbiting and 
exchanging regimes---may require special consideration when devising a wake detection protocol.

We note that the study here demonstrates the viability of using hydrodynamic 
signals to detect and classify wakes in an idealized setting;
 much work remains to be done to demonstrate the viability of doing the same 
in the face of numerous practical real-world challenges.
In the present study, 
numerous constraints were imposed---both on the wake parameters and on the 
relative location and orientation of the body-wake system---in order 
to demonstrate the viability of classifying wakes from 
their hydrodynamic signatures.
In future investigations, wake classification libraries will need to be 
constructed with a richer set of entries that incorporate factors such 
as relative locations and alignment of the wake and body.
Further, only velocity measurements taken at 
a single point on the body were considered here, 
which could limit the ability of wake detection 
protocol to distinguish between wake generating systems.
For example, at a sufficient distance, 
a 2P wake and 2S wake will impart 
strikingly similar hydrodynamic signatures.
Additionally, sensor noise and low 
signal-to-noise ratios
will certainly play an important role in performance as must be addressed in the future, 
perhaps by means of more sophisticated 
signal processing techniques.

In addressing these practical challenges, we can 
draw inspiration from biology once again.
For example, the lateral line system is composed of arrays of 
both superficial and canal neuromasts, giving marine swimmers access to a multi-modal and distributed sensing system.
Indeed, multi-modal distributed sensing 
will be more likely to succeed at distinguishing 
between wakes when body-wake alignment 
and separation are unknown and need to be accounted for.
Moreover, measurements of pressure and pressure gradients---as in canal neuromasts---may serve to amplify hydrodynamic signals and yield improved performance in low signal-to-noise environments.
It may also be possible to devise swimming gaits that are optimized for wake sensing and detection.
Recent studies of biological swimmers show that properly timed head motions can lead to a threefold
sensitivity of the lateral line system without compromising other performance objectives, such as propulsive efficiency~\cite{akanyetiNC2016}.
%
Certainly, distributed and multi-modal sensing combined 
with sensing-optimized gaits will play a role in making 
wake detection protocols more practically relevant for use in
bioinspired robotic swimmers in the future.

On a final note, the results of the present study suggest that
wake dynamics are important to the ability distinguish between exotic wakes.
Thus, it stands that wake dynamics and regimes of motion must be taken into 
consideration in future studies of hydrodynamic sensing and locomotion in 
biological and bioinspired systems alike.

\section*{Acknowledgments}
The authors gratefully acknowledge support from the Department of Aerospace Engineering and 
Mechanics at the Universityof Minnesota.

\nocite{}
\bibliographystyle{iopart-num}
\bibliography{main}

\providecommand{\newblock}{}
\begin{thebibliography}{10}
\expandafter\ifx\csname url\endcsname\relax
  \def\url#1{{\tt #1}}\fi
\expandafter\ifx\csname urlprefix\endcsname\relax\def\urlprefix{URL }\fi
\providecommand{\eprint}[2][]{\url{#2}}

\bibitem{arnoldBR1974}
Arnold G~P 1974 {\em Biological Reviews\/} {\bf 49} 515--576 ISSN 1469-185X
  \urlprefix\url{http://dx.doi.org/10.1111/j.1469-185X.1974.tb01173.x}

\bibitem{liaoSci2003}
Liao J~C, Beal D~N, Lauder G~V and Triantafyllou M~S 2003 {\em Science\/} {\bf
  302} 1566--1569

\bibitem{marrasBES2015}
Marras S, Killen S~S, Lindstr\"om J, McKenzie D~J, Steffensen J~F and Domenici
  P 2015 {\em Behavioral Ecology and Sociobiology\/} {\bf 69} 219--226

\bibitem{pitcherSci1976}
Pitcher T~J, Partridge B~L and Wardle C~S 1976 {\em Science (New York, N.Y.)\/}
  {\bf 194} 963--965 ISSN 0036-8075

\bibitem{montgomeryEBF2001}
Montgomery J~C, Coombs S and Baker C~F 2001 {\em Environmental Biology of
  Fishes\/} {\bf 62} 87--96 ISSN 1573-5133
  \urlprefix\url{https://doi.org/10.1023/A:1011873111454}

\bibitem{windsorJEB2010_1}
Windsor S~P, Norris S~E, Cameron S~M, Mallinson G~D and Montgomery J~C 2010
  {\em Journal of Experimental Biology\/} {\bf 213} 3819--3831 ISSN 0022-0949
  (\textit{Preprint}
  \eprint{http://jeb.biologists.org/content/213/22/3819.full.pdf})
  \urlprefix\url{http://jeb.biologists.org/content/213/22/3819}

\bibitem{windsorJEB2010_2}
Windsor S~P, Norris S~E, Cameron S~M, Mallinson G~D and Montgomery J~C 2010
  {\em Journal of Experimental Biology\/} {\bf 213} 3832--3842 ISSN 0022-0949
  (\textit{Preprint}
  \eprint{http://jeb.biologists.org/content/213/22/3832.full.pdf})
  \urlprefix\url{http://jeb.biologists.org/content/213/22/3832}

\bibitem{bleckmann1994reception}
Bleckmann H 1994 {\em Reception of hydrodynamic stimuli in aquatic and
  semiaquatic animals\/} ({\em Progress in Zoology\/} vol~41) (New York: Gustav
  Fischer Verlag)

\bibitem{speddingAnnRev2014}
Spedding G~R 2014 {\em Annual Review of Fluid Mechanics\/} {\bf 46} 273--302

\bibitem{triantafyllouAnnRev2016}
Triantafyllou M~S, Weymouth G~D and Miao J 2016 {\em Annual Review of Fluid
  Mechanics\/} {\bf 48} 150720185944007 ISSN 0066-4189
  \urlprefix\url{http://www.annualreviews.org/doi/abs/10.1146/annurev-fluid-122414-034329}

\bibitem{bleckmann2009lateral}
Bleckmann H and Zelick R 2009 {\em Integrative zoology\/} {\bf 4} 13--25

\bibitem{weihsJTB1984}
Weihs D and Webb P 1984 {\em Journal of Theoretical Biology\/} {\bf 106} 189 --
  206 ISSN 0022-5193
  \urlprefix\url{http://www.sciencedirect.com/science/article/pii/0022519384900195}

\bibitem{gemmellJRSI2014}
Gemmell B~J, Adhikari D and Longmire E~K 2014 {\em Journal of the Royal Society
  Interface\/} {\bf 11} 20130880
  \urlprefix\url{http://www.ncbi.nlm.nih.gov/pmc/articles/PMC3836328/}

\bibitem{pohlmannPNAS2001}
Pohlmann K, Grasso F~W and Breithaupt T 2001 {\em Proceedings of the National
  Academy of Sciences\/} {\bf 98} 7371--7374 (\textit{Preprint}
  \eprint{http://www.pnas.org/content/98/13/7371.full.pdf})
  \urlprefix\url{http://www.pnas.org/content/98/13/7371.abstract}

\bibitem{sichert2009hydrodynamic}
Sichert A~B, Bamler R and van Hemmen J~L 2009 {\em Physical review letters\/}
  {\bf 102} 058104

\bibitem{bouffanais2010hydrodynamic}
Bouffanais R, Weymouth G~D and Yue D~K 2010 Hydrodynamic object recognition
  using pressure sensing {\em Proceedings of the Royal Society of London A:
  Mathematical, Physical and Engineering Sciences\/} (The Royal Society) p
  rspa20100095

\bibitem{chambers2014fish}
Chambers L, Akanyeti O, Venturelli R, Je{\v{z}}ov J, Brown J, Kruusmaa M,
  Fiorini P and Megill W 2014 {\em Journal of The Royal Society Interface\/}
  {\bf 11} 20140467

\bibitem{kleinBJN2011}
Klein A and Bleckmann H 2011 {\em Beilstein Journal of Nanotechnology\/} {\bf
  2} 276--283 ISSN 21904286

\bibitem{franoschPRL2009}
Franosch J~M~P, Hagedorn H~J~A, Goulet J, Engelmann J and {Van Hemmen} J~L 2009
  {\em Physical Review Letters\/} {\bf 103} 1--4 ISSN 00319007

\bibitem{renBB2012}
Ren Z and Mohseni K 2012 {\em Bioinspiration {\&} Biomimetics\/} {\bf 7} 036016
  ISSN 1748-3182

\bibitem{devriesBB2015}
DeVries L, Lagor F~D, Lei H, Tan X and Paley D 2015 {\em Bioinspiration {\&}
  Biomimetics\/} {\bf 10} 025002 ISSN 1748-3190

\bibitem{colvertJFM2016}
Colvert B and Kanso E 2016 {\em Journal of Fluid Mechanics\/} {\bf 793}
  656--666 ISSN 0022-1120
  \urlprefix\url{http://www.journals.cambridge.org/abstract{\_}S0022112016001415}

\bibitem{borazjaniJEB2008}
Borazjani I and Sotiropoulos F 2008 {\em Journal of Experimental Biology\/}
  {\bf 211} 1541--1558

\bibitem{williamsonJFS1988}
Williamson C and Roshko A 1988 {\em Journal of fluids and structures\/} {\bf 2}
  355--381

\bibitem{arefJFS2006}
Aref H, Stremler M~A and Ponta F~L 2006 {\em Journal of Fluids and
  Structures\/} {\bf 22} 929--940 ISSN 08899746

\bibitem{kern2006simulations}
Kern S and Koumoutsakos P 2006 {\em Journal of Experimental Biology\/} {\bf
  209} 4841--4857

\bibitem{lentinkJEB2008}
Lentink D, Muijres F~T, Donker-Duyvis F~J and van Leeuwen J~L 2008 {\em Journal
  of Experimental Biology\/} {\bf 211} 267--273 ISSN 0022-0949

\bibitem{schnipperJFM2009}
Schnipper T, Andersen A and Bohr T 2009 {\em Journal of Fluid Mechanics\/} {\bf
  633} 411--423

\bibitem{mooredJFM2012}
Moored K~W, Dewey P~A, Smits A~J and {Haj-Hariri} H 2012 {\em Journal of Fluid
  Mechanics\/} {\bf 708} 329--348

\bibitem{smits2014}
Smits A~J, Moored K~W and Dewey P~A 2014 {\em Fluid-Structure-Sound
  Interactions and Control: Proceedings of the 2nd Symposium on
  Fluid-Structure-Sound Interactions and Control\/} (Berlin, Heidelberg:
  Springer Berlin Heidelberg) chap The Swimming of Manta Rays, pp 291--300 ISBN
  978-3-642-40371-2
  \urlprefix\url{http://dx.doi.org/10.1007/978-3-642-40371-2_43}

\bibitem{tytellICB2010}
Tytell E~D, Borazjani I, Sotiropoulos F, Baker T~V, Anderson E~J and Lauder G~V
  2010 {\em Integrative and comparative biology\/} {\bf 50} 1140--1154

\bibitem{basuPOF2015}
Basu S and Stremler M~A 2015 {\em Physics of Fluids\/} {\bf 27} 103603

\bibitem{stremlerTCFD2010}
Stremler M~A 2010 {\em Theoretical and Computational Fluid Dynamics\/} {\bf 24}
  25--37

\bibitem{stremlerJFS2011}
Stremler M~A, Salmanzadeh A, Basu S and Williamson C~H~K 2011 {\em Journal of
  Fluids and Structures\/} {\bf 27} 774--783 ISSN 08899746

\bibitem{mullerJEB2008}
M{\"{u}}ller U~K, van~den Boogaart J~G~M and van Leeuwen J~L 2008 {\em Journal
  of Experimental Biology\/} {\bf 211} 196--205 ISSN 0022-0949

\bibitem{hematiAIAA2016}
Hemati M~S 2016 Learning wake regimes from snapshot data {\em AIAA Paper
  2016-3781\/} (Washington, DC: 46th AIAA Fluid Dynamics Conference, AIAA
  Aviation)

\bibitem{bishop2006}
Bishop C~M 2006 {\em {Pattern Recognition and Machine Learning}\/} (Springer)
  ISBN 9780387310732

\bibitem{tan2006}
Tan P~N, Steinbach M and Kumar V 2006 {\em Introduction to Data Mining\/}
  (Boston: Pearson)

\bibitem{newton2001}
Newton P~K 2001 {\em The $N$-Vortex Problem: Analytical Techniques\/} (New
  York: Springer-Verlag)

\bibitem{cottet2000}
Cottet G~H and Koumoutsakos P~D 2000 {\em Vortex Methods: Theory and
  Practice\/} (New York: Cambridge University Press)

\bibitem{katz2001}
Katz J and Plotkin A 2001 {\em Low-Speed Aerodynamics\/} 2nd ed (New York:
  Cambridge University Press)

\bibitem{akanyetiNC2016}
Akanyeti O, Thornycroft P~J~M, Lauder G~V, Yanagitsuru Y~R, Peterson A~N and
  Liao J~C 2016 {\em Nature Communications\/} {\bf 7} 11044 ISSN 2041-1723
  \urlprefix\url{http://www.nature.com/doifinder/10.1038/ncomms11044}

\end{thebibliography}

\end{document}